\documentclass[aps,prl,twocolumn,superscriptaddress]{revtex4-1}

\usepackage{graphicx,latexsym}
\usepackage{amsmath,amssymb,amsfonts}
\usepackage{bm}
\usepackage{braket}

\begin{document}


\title{Theory of Non-Hermitian Fermionic Superfluidity with a Complex-Valued Interaction}
\author{Kazuki Yamamoto}
\email{yamamoto.kazuki.72n@st.kyoto-u.ac.jp}
\affiliation{Department of Physics, Kyoto University, Kyoto 606-8502, Japan}
\author{Masaya Nakagawa}
\affiliation{RIKEN Center for Emergent Matter Science (CEMS), Wako, Saitama 351-0198, Japan}
\affiliation{Department of Physics, University of Tokyo, 7-3-1 Hongo, Bunkyo-ku, Tokyo 113-0033, Japan}
\author{Kyosuke Adachi}
\affiliation{Department of Physics, Kyoto University, Kyoto 606-8502, Japan}
\affiliation{RIKEN Center for Biosystems Dynamics Research (BDR), Kobe, Hyogo, 650-0047, Japan}
\author{Kazuaki Takasan}
\affiliation{Department of Physics, Kyoto University, Kyoto 606-8502, Japan}
\affiliation{Department of Physics, University of California, Berkeley, CA 94720, USA}
\author{Masahito Ueda}
\affiliation{Department of Physics, University of Tokyo, 7-3-1 Hongo, Bunkyo-ku, Tokyo 113-0033, Japan}
\affiliation{RIKEN Center for Emergent Matter Science (CEMS), Wako, Saitama 351-0198, Japan}
\author{Norio Kawakami}
\affiliation{Department of Physics, Kyoto University, Kyoto 606-8502, Japan}




\date{\today}

\begin{abstract}
Motivated by recent experimental advances in ultracold atoms, we analyze a non-Hermitian (NH) BCS Hamiltonian with a complex-valued interaction arising from inelastic scattering between fermions. We develop a mean-field theory to obtain a NH gap equation for order parameters, which are different from the standard BCS ones due to the inequivalence of left and right eigenstates in the NH physics. We find unconventional phase transitions unique to NH systems: superfluidity shows reentrant behavior with increasing dissipation, as a consequence of non-diagonalizable exceptional points, lines, and surfaces in the quasiparticle Hamiltonian for weak attractive interactions. For strong attractive interactions, the superfluid gap never collapses but is enhanced by dissipation due to an interplay between the BCS-BEC crossover and the quantum Zeno effect. Our results lay the groundwork for studies of fermionic superfluidity subject to inelastic collisions.
\end{abstract}



\maketitle


\textit{Introduction}.--\
In recent years, non-Hermitian (NH) quantum systems have been actively studied both experimentally and theoretically \cite{Bender98, Bender07, Nakagawa18, Lourenco18, Ashida16, Ashida17, Kawabata17, Rudner09, Lee16, Shen18, Kunst18, Yao18, Kawabata18, Gong18, Kawabata19, Kawabata18a, Lee18, Lee14, Ghatak18, Zhou18a, Chtchelkatchev12, Guo09, Ruter10, Feng11, Regensburger12, Zeuner15, Li19, Xiao17, Zhou18, Xiao18, Durr09, Garcia09, Daley09, Kantian09, Yoshida18, Zhong18}. NH quantum systems arise when the system undergoes dissipation to an environment \cite{Daley14, Dalibard92}. It has been revealed that non-Hermiticity drastically alters the properties of a number of quantum phenomena that have been established in the Hermitian physics, such as quantum phase transitions \cite{Bender98, Bender07, Nakagawa18, Lourenco18}, quantum critical behavior \cite{Ashida16, Ashida17, Kawabata17}, topological phases \cite{Rudner09, Lee16, Shen18, Kunst18, Yao18, Kawabata18, Gong18, Kawabata19, Kawabata18a, Lee18}, and magnetism \cite{Lee14}. Such theoretical predictions have been confirmed experimentally by using optical systems and ultracold atoms \cite{Guo09, Ruter10, Feng11, Regensburger12, Zeuner15, Li19, Xiao17, Zhou18, Xiao18}. However, since most of the previous studies dealt with single-particle physics, exploration of many-body physics in NH systems is still in its infancy \cite{Nakagawa18, Lourenco18, Ashida16, Ashida17, Durr09, Garcia09, Daley09, Kantian09}.

Fermionic superfuidity is one of the most striking quantum many-body phenomena, which has been a subject of intensive investigation in condensed matter physics \cite{Bardeen57}. More recently, ultracold atomic systems have opened a new arena to study fermionic superfluidity \cite{Regal04, Zwierlein04, Bartenstein04}, where they are subject to losses due to inelastic collisions. For example, if we consider a superfluid mediated by the orbital Feshbach resonance \cite{Zhang15, Iskin16, Xu16, He16}, which controls an interaction between the ground state and an excited state of an atom \cite{Hofer15, Pagano15, Cappellini19, Folling19}, loss inevitably occurs due to inelastic processes between different orbitals. Such inelastic two-body losses cause the decay of eigenstates of the Hamiltonian and may be described by complex-valued interactions, thus providing an ideal platform to study NH fermionic superfluids. Despite its growing importance, however, theory for NH fermionic superfluidity has not been established yet \cite{Ghatak18, Zhou18a, Chtchelkatchev12}.

In this Letter, we demonstrate how fermionic superfluidity in ultracold atoms is modified under inelastic collisions, by generalizing the standard BCS theory to a situation in which fermions interact with each other via a complex-valued attraction. We elucidate that the non-Hermiticity alters several fundamental properties of superfluidity; for example, the order parameters of particles and holes are not necessarily complex conjugate to each other in the NH physics, and the Bogoliubov quasiparticles obey neither Fermi nor Bose statistics since eigenstates are, in general, not orthogonal to each other.

Furthermore, we find that the non-Hermiticity leads to unique quantum phase transitions in superfluids. For a weak interaction, the real part of the superfluid gap is first suppressed and then quenched with increasing dissipation. Remarkably, superfluidity is restored beyond a certain strength of dissipation and the superfluid gap is even enhanced afterwards with increasing dissipation. We show that these phase transitions emerge from exceptional  points, lines and surfaces that are unique to the NH physics, where the Hamiltonian cannot be diagonalized \cite{Heiss12, Berry04}. For a strong interaction, superfluidity is not suppressed and never breaks down because fermions are paired to form molecules on each site, thereby avoiding intersite decoherence. Our finding can experimentally be tested in various ultracold atomic species under inelastic collisions such as $^{173}$Yb, $^{40}$K, and $^{6}$Li \cite{Cappellini19, Folling19, Stewart08, Chin04, Ketterle06, Esslinger06, Bakr18}.


\textit{Model}.--\
We consider ultracold fermionic atoms with an attractive interaction in a three-dimensional optical lattice. When atoms undergo inelastic collisions, the scattered atoms are lost from the system since a large internal energy is converted to the kinetic energy. An atomic gas undergoing two-body losses due to inelastic collisions is described by a quantum master equation \cite{Daley14}
\begin{align}
\dot{\rho}
&=-\frac{i}{\hbar}[H,\rho]-\frac{1}{2}\gamma\sum_i\left(L_i^\dagger{L}_i\rho+\rho{L}_i^\dagger{L}_i-2L_i\rho{L}_i^\dagger\right)\notag\displaybreak[2]\\
&=-i(H_{\mathrm{eff}}\rho-\rho{H}_{\mathrm{eff}}^\dagger)+\gamma\sum_i{L}_i\rho{L}_i^\dagger,
\label{eq_master}
\end{align}
where $L_i$ is a Lindblad operator that describes a loss at site $i$ with rate $\gamma$, and $\rho$ is the density matrix of the atomic gas. When the quantum-jump term, which is the last term in Eq.\ \eqref{eq_master}, is negligible, the system is described by an effective NH Hamiltonian $H_{\mathrm{eff}}=H-\frac{i}{2}\gamma\sum_iL_i^\dagger{L}_i$. Such a situation is realized when we consider the dynamics over a sufficiently short time compared with the inverse loss rate $1/\gamma$ \cite{Durr09}, which characterizes the timescale where the effect of quantum jumps becomes significant. In this case, the lowest real part of the eigenspectrum gives the effective ground state, and the imaginary part of energy corresponds to a decay rate of each eigenstate. The two-body loss is described by $L_i=c_{i\downarrow}c_{i\uparrow}$, giving a NH BCS Hamiltonian
\begin{align}
H_{\mathrm{eff}}=\sum_{\bm{k}\sigma}\xi_{\bm{k}}c_{\bm{k}\sigma}^\dagger{c}_{\bm{k}\sigma}-U\sum_{i}c_{i\uparrow}^{\dagger}c_{i\downarrow}^{\dagger}c_{i\downarrow}c_{i\uparrow},
\label{eq_Hubbard}
\end{align}
with a \textit{complex-valued} interaction $U=U_1+i\gamma/2$, where $U_1,\gamma>0$. Here, $\xi_{\bm{k}}\equiv\epsilon_{\bm{k}}-\mu$, $\epsilon_{\bm{k}}$ is the energy dispersion, $\mu$ is the chemical potential, and $c_{\bm{k}\sigma}$ and $c_{i\sigma}$ denote annihilation operators of a spin-$\sigma$ fermion with momentum $\bm{k}$ and at site $i$, respectively. In this Letter, we formulate a mean-field theory from $H_{\mathrm{eff}}$ and elucidate how unconventional properties of superfluidity emerge in NH BCS systems.


\textit{Formulation of the NH mean-field theory}.--\ 
We first clarify how the standard BCS mean-field theory is changed due to non-Hermiticity by formulating it with a path-integral approach. We start with a partition function defined as
\begin{align}
Z=\sum_ne^{-\beta{E}_n}=\sum_n~_L\langle E_n|e^{-\beta{H_{\mathrm{eff}}}}|E_n\rangle_R.
\label{eq_partitionfunction}
\end{align}
Here, $_L\langle E_n|$ and $|E_n\rangle_R$ are left and right eigenstates of $H_{\mathrm{eff}}$ with eigenenergy $E_n$, and they satisfy an orthonormal relation $_L\langle E_n|E_m\rangle_R=\delta_{nm}$ \cite{Brody13}. We note that, as temperature is not well defined in generic open quantum systems, we only consider the infinite limit of $\beta$ to elucidate the physics of the ground state and calculate the excitation spectrum. Thus, $\beta$ is a parameter used to formulate a path integral and should not be regarded as the temperature of the system. We use a path-integral representation of the partition function \eqref{eq_partitionfunction} to perform the Hubbard-Stratonovich transformation with auxiliary fields $\Delta$, $\bar{\Delta}$ and then integrate out the fermionic degrees of freedom to obtain $Z=\int\mathcal{D}\bar{\Delta}\mathcal{D}{\Delta}e^{-S_{\mathrm{eff}}(\Delta,\bar{\Delta})}$, where $S_{\mathrm{eff}}$ is the effective action given by \cite{Supple}
\begin{align}
S_\mathrm{eff}(\bar{\Delta},\Delta)
=-\sum_{\omega_n,\bm{k}}\log(\omega_n^2+\xi_{\bm{k}}^2+\bar{\Delta}{\Delta})+\frac{\beta{N}}{U}\bar{\Delta}{\Delta}.
\label{eq_actioneff}
\end{align}
Here, $N$ denotes the number of lattice sites, and $\omega_n$ is the Matsubara frequency of fermions. The saddle point condition for the partition function, $\partial{S}_\mathrm{eff}/\partial\Delta=\partial{S}_\mathrm{eff}/\partial\bar{\Delta}=0$, yields the NH gap equation
\begin{align}
\frac{N}{U}=\sum_{\bm{k}}\frac{1}{2\sqrt{\xi_{\bm{k}}^2+\bar{\Delta}\Delta}}\tanh\frac{\beta\sqrt{\xi_{\bm{k}}^2+\bar{\Delta}\Delta}}{2},
\label{eq_gap}
\end{align}
if there exists a nontrivial solution other than $\Delta=\bar{\Delta}=0$, where we set $\beta$ to infinity to obtain an effective ground state. The chemical potential $\mu$ is determined so that the mean particle number in the non-Hermitian ensemble \eqref{eq_partitionfunction} is equal to the particle number of the density-matrix sector of interest \cite{Supple}.

In NH physics, four distinct types of order parameters can be defined according to whether left and right eigenstates are assigned to the bra or ket vectors in the expectation value. Importantly, the expectation value of an operator $A$ calculated from the non-Hermitian ensemble \eqref{eq_partitionfunction} should correspond to $~_L\langle{A}\rangle_R\equiv\sum_n~_L\langle E_n|A|E_n\rangle_Re^{-\beta{E}_n}/Z$. Thus, the order parameters corresponding to the superfluid gap are given by \cite{Supple}
\begin{gather}
\Delta=-\frac{U}{N}\sum_{\bm{k}}{}_L\langle c_{-\bm{k}\downarrow}c_{\bm{k}\uparrow}\rangle_R,\label{eq_OrderParameterA}\\
\bar{\Delta}=-\frac{U}{N}\sum_{\bm{k}}{}_L\langle c_{\bm{k}\uparrow}^\dagger{c}_{-\bm{k}\downarrow}^\dagger\rangle_R,
\label{eq_OrderParameterB}
\end{gather}
indicating that $\bar{\Delta}\neq\Delta^*$ since $|E_n\rangle_L\neq|E_n\rangle_R$. As discussed below, this leads to various intriguing consequences on the properties of a NH superfluid.

To elucidate the effect of non-Hermiticity, let us apply the mean-field decoupling to the NH BCS Hamiltonian with the simplest $s$-wave pairing interaction
\begin{align}
H_{\mathrm{eff}}=\sum_{\bm{k}\sigma}\xi_{\bm{k}}c_{\bm{k}\sigma}^\dagger{c}_{\bm{k}\sigma}-\frac{U}{N}\sum_{\bm{k}\bm{k}'}c_{\bm{k}\uparrow}^{\dagger}c_{-\bm{k}\downarrow}^\dagger{c}_{-\bm{k}'\downarrow}c_{\bm{k}'\uparrow}.
\label{eq_Hubbard_k}
\end{align}
Substituting $c_{\bm{k}\uparrow}^\dagger{c}_{-\bm{k}\downarrow}^\dagger={}_L\langle c_{\bm{k}\uparrow}^\dagger{c}_{-\bm{k}\downarrow}^\dagger\rangle_R+\delta({c}_{\bm{k}\uparrow}^\dagger{c}_{-\bm{k}\downarrow}^\dagger)$ and $c_{-\bm{k}\downarrow}c_{\bm{k}\uparrow}={}_L\langle c_{-\bm{k}\downarrow}c_{\bm{k}\uparrow}\rangle_R+\delta({c}_{-\bm{k}\downarrow}c_{\bm{k}\uparrow})$ into Eq.\ \eqref{eq_Hubbard_k} and neglecting the second-order terms in $\delta$, we obtain the mean-field Hamiltonian
\begin{align}
H_{\mathrm{MF}}
=\sum_{\bm{k}}
\left(
\begin{matrix}
c_{\bm{k}\uparrow}^\dagger&c_{-\bm{k}\downarrow}
\end{matrix}
\right)
\left(
\begin{matrix}
\xi_{\bm{k}}&\Delta\\
\bar{\Delta}&-\xi_{\bm{k}}
\end{matrix}
\right)
\left(
\begin{matrix}
c_{\bm{k}\uparrow}\\
c_{-\bm{k}\downarrow}^\dagger
\end{matrix}
\right),
\label{eq_meanfield}
\end{align}
which is diagonalized as $H_{\mathrm{MF}}=\sum_{\bm{k}}E_{\bm{k}}(\bar{\gamma}_{\bm{k}\uparrow}\gamma_{\bm{k}\uparrow}+\bar{\gamma}_{-\bm{k}\downarrow}\gamma_{-\bm{k}\downarrow})-\sum_{\bm{k}}E_{\bm{k}}$. The quasiparticle operators $\bar{\gamma}_{\bm{k}\sigma}$, $\gamma_{\bm{k}\sigma}$ and the corresponding energy $E_{\bm{k}}$ are given by $\bar{\gamma}_{\bm{k}\uparrow}=u_{\bm{k}}c_{\bm{k}\uparrow}^\dagger-\bar{v}_{\bm{k}}{c}_{-\bm{k}\downarrow}$, $\gamma_{\bm{k}\uparrow}=u_{\bm{k}}c_{\bm{k}\uparrow}-v_{\bm{k}}c_{-\bm{k}\downarrow}^\dagger$ (and similar equations hold for $\bar{\gamma}_{\bm{k}\downarrow}$, $\gamma_{\bm{k}\downarrow}$) and $E_{\bm{k}}=\sqrt{\xi_{\bm{k}}^2+\bar{\Delta}\Delta}$, respectively, where the coefficients satisfy $u_{\bm{k}}^2+v_{\bm{k}}\bar{v}_{\bm{k}}=1$ (for their explicit forms, see Supplemental Material \cite{Supple}). In the Hermitian limit, $\bar{\gamma}_{\bm{k}\sigma}$ and $\bar v_{\bm k}$ respectively reduce to $\gamma^\dagger_{\bm{k}\sigma}$ and $v^*_{\bm{k}}$ which describe the Bogoliubov quaiparticles. Here, we note that the mean-field Hamiltonian is non-Hermitian since $\Delta^*\neq\bar{\Delta}$, and the quasiparticle operators $\gamma_{\bm{k}\sigma}$ and $\bar{\gamma}_{\bm{k}\sigma}$ are not Hermitian conjugate to each other. Therefore, the Hamiltonian cannot be diagonalized via a unitary transformation.
\begin{figure}[t]
\includegraphics[width=8.5cm]{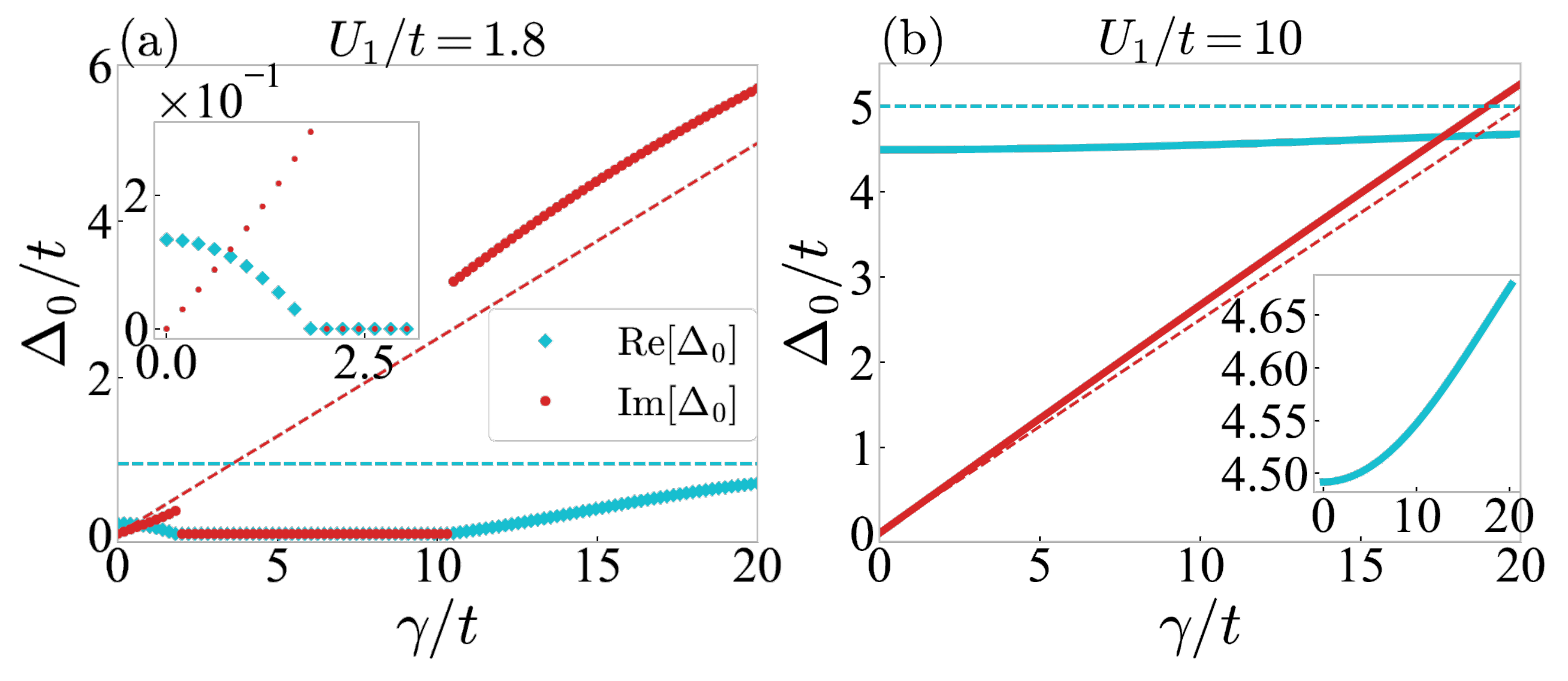}
\caption{Numerical solution of $\Delta_0$ as a function of $\gamma/t$ obtained from the NH gap equation \eqref{eq_gap} at $\beta\to\infty$ and $\mu=0$ for (a) $U_1/t=1.8$ and (b) $U_1/t=10$. We assume a cubic lattice with energy dispersion $\epsilon_{\bm{k}}=-2t(\cos{k_x}+\cos{k_y}+\cos{k_z})$, where $t$ is the hopping amplitude. The dashed lines denote the asymptotic behavior in the strong-dissipation limit. The insets show (a) an enlarged view near the origin (weak dissipation) and (b) that of the real part.}
\label{fig_gapN}
\end{figure}

As a result, the right and left ground states are defined by $\gamma_{\bm{k}\sigma}|\mathrm{BCS}\rangle_R=0$ and $\bar{\gamma}_{\bm{k}\sigma}^\dagger|\mathrm{BCS}\rangle_L=0$, respectively, where
\begin{align}
&|\mathrm{BCS}\rangle_R=\prod_{\bm{k}}\left(u_{\bm{k}}+v_{\bm{k}}c_{\bm{k}\uparrow}^\dagger{c}_{-\bm{k}\downarrow}^\dagger\right)\ket{0},\label{eq_BCSR}\\
&|\mathrm{BCS}\rangle_L=\prod_{\bm{k}}\left(u_{\bm{k}}^*+\bar{v}_{\bm{k}}^*c_{\bm{k}\uparrow}^\dagger{c}_{-\bm{k}\downarrow}^\dagger\right)\ket{0},
\label{eq_BCSL}
\end{align}
and $\ket{0}$ is the vacuum for fermions. They satisfy $_L\langle\mathrm{BCS}|\mathrm{BCS}\rangle_R=1$ and reproduce the ordinary BCS ground state in the Hermitian limit. We thus obtain $H\bar{\gamma}_{\bm{k}\sigma}|\mathrm{BCS}\rangle_R=E_{\bm{k}}\bar{\gamma}_{\bm{k}\sigma}\mathrm{BCS}\rangle_R$ and $H^\dagger\gamma_{\bm{k}\sigma}^\dagger|\mathrm{BCS}\rangle_L=E_{\bm{k}}^*\gamma_{\bm{k}\sigma}^\dagger|\mathrm{BCS}\rangle_L$, which imply that $\bar{\gamma}_{\bm{k}\sigma}$ and $\gamma_{\bm{k}\sigma}^\dagger$ create the right and left eigenstates, respectively, when acted on the ground state. Here, we have shifted the ground state energy to zero. Using Eqs.\ \eqref{eq_OrderParameterA}, \eqref{eq_OrderParameterB}, \eqref{eq_BCSR}, and \eqref{eq_BCSL}, we obtain the $\beta\to\infty$ limit of the NH gap equation as $N/U=\sum_{\bm{k}}1/2E_{\bm{k}}$, which is solved self-consistently. We note that the quasiparticle operators satisfy an anticommutation relation $\{\gamma_{\bm{k}\sigma},\bar{\gamma}_{\bm{k}'\sigma'}\}=\delta_{\bm{k}\bm{k}'}\delta_{\sigma\sigma'}$, although these quasiparticles obey neither Fermi nor Bose statistics due to $\gamma_{\bm{k}\sigma}^\dag\neq\bar{\gamma}_{\bm{k}\sigma}$, reflecting non-Hermiticity of the mean-field Hamiltonian.

Here, we point out an important relation between the order parameters $\Delta$ and $\bar{\Delta}$. In the Hermitian case, they are complex conjugate to each other and we can choose a gauge where $\Delta$ is real without loss of generality. This is equivalent to requiring $H_{\mathrm{MF}}^\dagger=H_{\mathrm{MF}}^*$ in the matrix representation in the Fock-state basis in the Hilbert space. Now we consider the NH case. As in the Hermitian case, the NH BCS Hamiltonian \eqref{eq_Hubbard_k} satisfies a symmetry relation $H^\dag=H^\ast$ under the matrix representation in terms of Fock states, indicating that the left eigenstates are obtained through complex conjugation of the right ones. The NH BCS Hamiltonian has the U(1) symmetry as in the Hermitian case and this is not affected by the complex nature of the interaction. Then, when a superfluid is formed, its ground states become degenerate due to spontaneous U(1) symmetry breaking. The BCS ground states \eqref{eq_BCSR} and \eqref{eq_BCSL} are consistent with these properties if
\begin{align}
&\Delta(\theta)=\Delta_0e^{i\theta},\notag\\
&\bar{\Delta}(\theta)=\Delta_0e^{-i\theta},
\label{eq_OP}
\end{align}
where $\Delta_0\in\mathbb{C}$ and $\theta$ is the U(1) phase. By choosing a special gauge for which $H_{\mathrm{MF}}^\dagger=H_{\mathrm{MF}}^*$ is satisfied, we have $\Delta=\bar{\Delta}$. Here, we note that the relation \eqref{eq_OP} is specific to the NH BCS Hamiltonian \eqref{eq_Hubbard_k} and may be changed depending on symmetry of a NH Hamiltonian.
\begin{figure}[t]
\includegraphics[width=8.5cm]{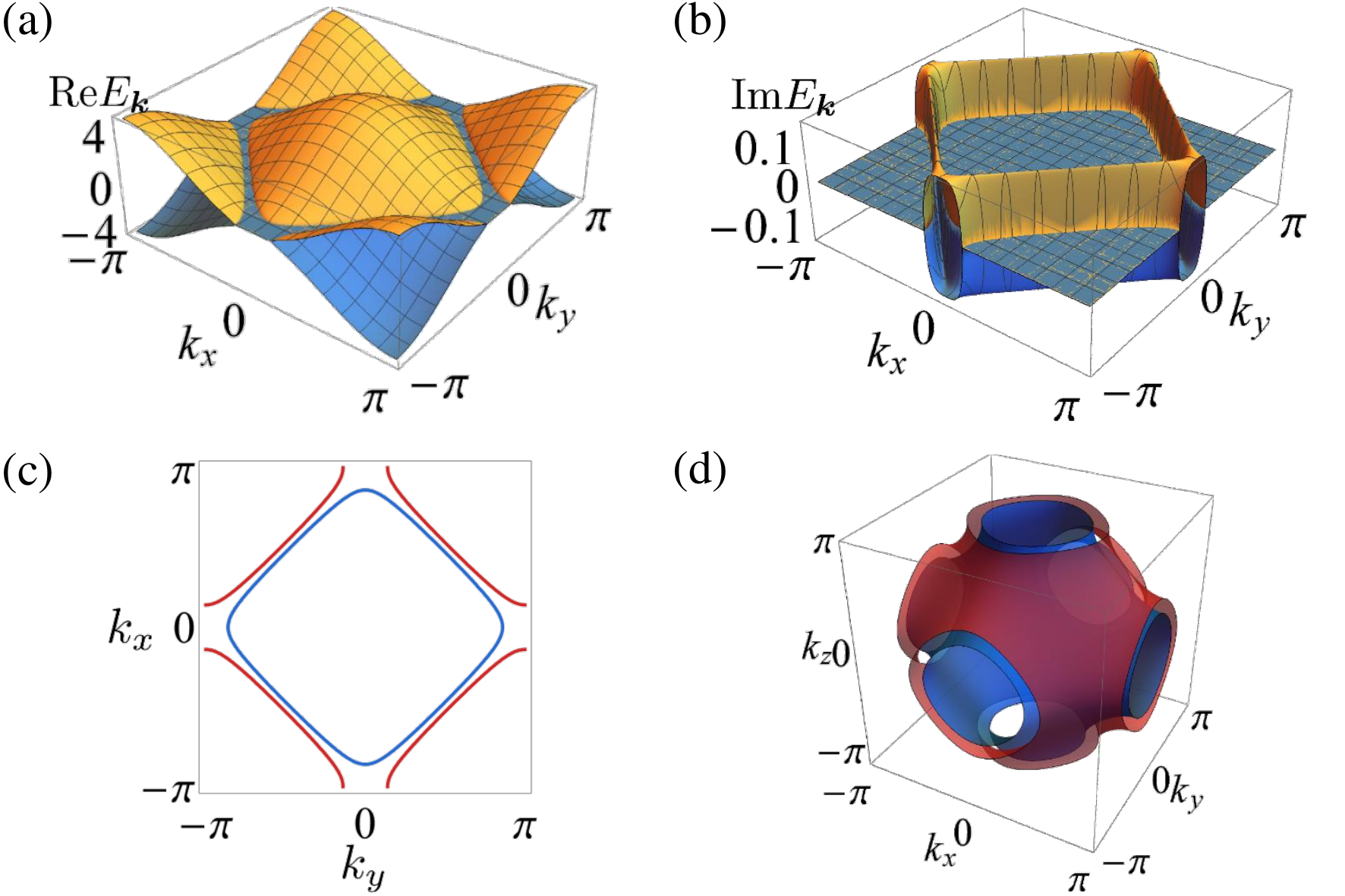}
\caption{(a) Real and (b) imaginary parts of the quasiparticle energy spectrum $E_{\bm{k}}=\pm\sqrt{\epsilon_{\bm{k}}^2+\Delta_0^2}$ (orange for positive sign, blue for negative sign) at critical points. (c) Exceptional lines in two dimensions. (d) Exceptional surfaces in three dimensions. The energy dispersion is for a square lattice $\epsilon_{\bm{k}}=-2(\cos{k_x}+\cos{k_y})$ in (a), (b) and (c), and for a cubic lattice $\epsilon_{\bm{k}}=-2(\cos{k_x}+\cos{k_y}+\cos{k_z})$ in (d). The gap is set to $\Delta_0=0.19i$ for (a), (b) and (c), and $\Delta_0=0.4i$ for (d).}
\label{fig_exceptional}
\end{figure}


\textit{Quantum phase transitions of a NH superfluid}.--\
We solve the gap equation at $\beta\to\infty$ numerically (for analytic solutions in the case of constant density of states, see Supplemental Material \cite{Supple}). Figure \ref{fig_gapN} shows the superfluid order parameter $\Delta_0$. Here, for simplicity, we consider a system with particle-hole symmetry, and set the chemical potential measured from the Fermi energy to zero \cite{Supple}. For small $U_1$, $\mathrm{Re}\Delta_0$ is suppressed by dissipation $\gamma$ and then vanishes, indicating a breakdown of superfluidity. Remarkably, as $\gamma$ increases, the superfluid solution reappears, and the gap size $\mathrm{Re}\Delta_0$ is enhanced due to dissipation, eventually exceeding the value in the Hermitian limit. On the other hand, for strong attractive interaction, $\mathrm{Re}\Delta_0$ is not suppressed, but rather enhanced due to dissipation.

The qualitative difference between the cases of weak and strong attractions is explained by an interplay between the BCS-BEC crossover \cite{Leggett80, Leggett80S, Nozieres85, Diener08} and dissipation. In the strong-dissipation limit, the behavior of the system is governed by the continuous quantum Zeno effect (QZE) \cite{Syassen08, Mark12, Barontini13, Zhu14, Tomita17, Yan13}, which suppresses tunneling to neighboring sites, leading to localization of particles. In the case of weak attraction, the inter-site coherence of  Cooper pairs is suppressed by dissipation and the superfluidity is destroyed. However, localization due to the QZE facilitates formation of on-site molecules of fermions for strong dissipation and consequently superfluidity reappears. In fact, the solution of the gap equation approaches $\Delta_0=U/2$ in the strong-dissipation limit, as shown by the dashed lines in Fig.~\ref{fig_gapN}. This is consistent with the fact that the physics is dominated by the on-site interaction under the QZE, supporting our mean-field analysis. On the other hand, under strong attraction, fermions form bosonic molecules almost at single sites and thus the molecules can survive under dissipation. In this case, the effect of dissipation is to give rise to an effective one-body loss of molecules and the remaining molecules can undergo Bose-Einstein condensation.

We here point out that the breakdown and restoration of superfluidity present clear signatures of the emergence of exceptional points, where the Hamiltonian cannot be diagonalized \cite{Heiss12, Berry04}. In fact, when $\mathrm{Re}\Delta_0=0$, the mean-field Hamiltonian $H_{\mathrm{MF}}$ cannot be diagonalized for $\xi_{\bm{k}}=\pm\mathrm{Im}\Delta_0$. Figure \ref{fig_exceptional}(a) shows the real part of the energy spectrum of quasiparticles in two dimensions. The regions where the orange and blue surfaces merge form exceptional points, lines (Fig.\ \ref{fig_exceptional}(c)), and surfaces (Fig.\ \ref{fig_exceptional}(d)) in one-, two-, and three-dimensional systems, respectively. Such characteristic behavior has its origin in a parity-particle-hole ($CP$) symmetry $CPH_{\mathrm{MF}}(\bm{k})(CP)^{-1}=-H_{\mathrm{MF}}(\bm{k})$ of the mean-field Hamiltonian $H_{\mathrm{MF}}(\bm{k})=\epsilon_{\bm{k}}\sigma_z+i\mathrm{Im}\Delta_0\sigma_x$, where $CP=\sigma_xK$, $\sigma_{x,z}$ are the Pauli matrices, and $K$ is complex conjugation \cite{Yoshida19, Budich19, Okugawa19, Zhou19}. In fact, as a function of the momentum, $H_{\mathrm{MF}}(\bm{k})$ exhibits spontaneous breaking of the $CP$ symmetry at the exceptional points, whose dimensionality is indeed protected by the symmetry constraint \cite{Supple}. Thus, the quantum phase transitions of the NH superfluid cannot be classified into the conventional first- or second-order phase transitions in Hermitian systems, but are attributed to the emergence of exceptional points unique to non-Hermiticity. We note here that the two exceptional points corresponding to the breakdown and restoration of superfluidity merge and disappear as the strength of attraction increases. Intriguingly, as a remnant of the merged exceptional points, the real part of the superfluid gap for intermediate strengths of $U_1$ shows a characteristic minimum at a certain strength of dissipation \cite{Supple}; the gap is first suppressed by dissipation, but enhanced again by the QZE as the dissipation is further increased.

Furthermore, the emergence of exceptional manifolds leads to some intriguing dynamics in the NH superfluid. In Fig.~\ref{fig_exceptional}(b), the imaginary part of the quasiparticle energy takes a positive finite value only in between the exceptional lines or surfaces, amplifying quasiparticle distribution in the particular region of the Brillouin zone through the time evolution. The characteristic structure in Fig.~\ref{fig_exceptional}(b), which can be used as a smoking gun of the non-Hermiticity, can be observed as long as $\mathrm{Im}\Delta_0>0$ even in a region away from the breakdown and restoration points.

We note that the nontrivial solution of the NH gap equation may give a metastable superfluid, which corresponds to a local minimum of the real part of the energy. Whether the superfluid is metastable or not is decided from comparison of ground-state energies between the superfluid state and the normal state, as detailed in Supplemental Material \cite{Supple}.
\begin{figure}
\includegraphics[width=8.5cm]{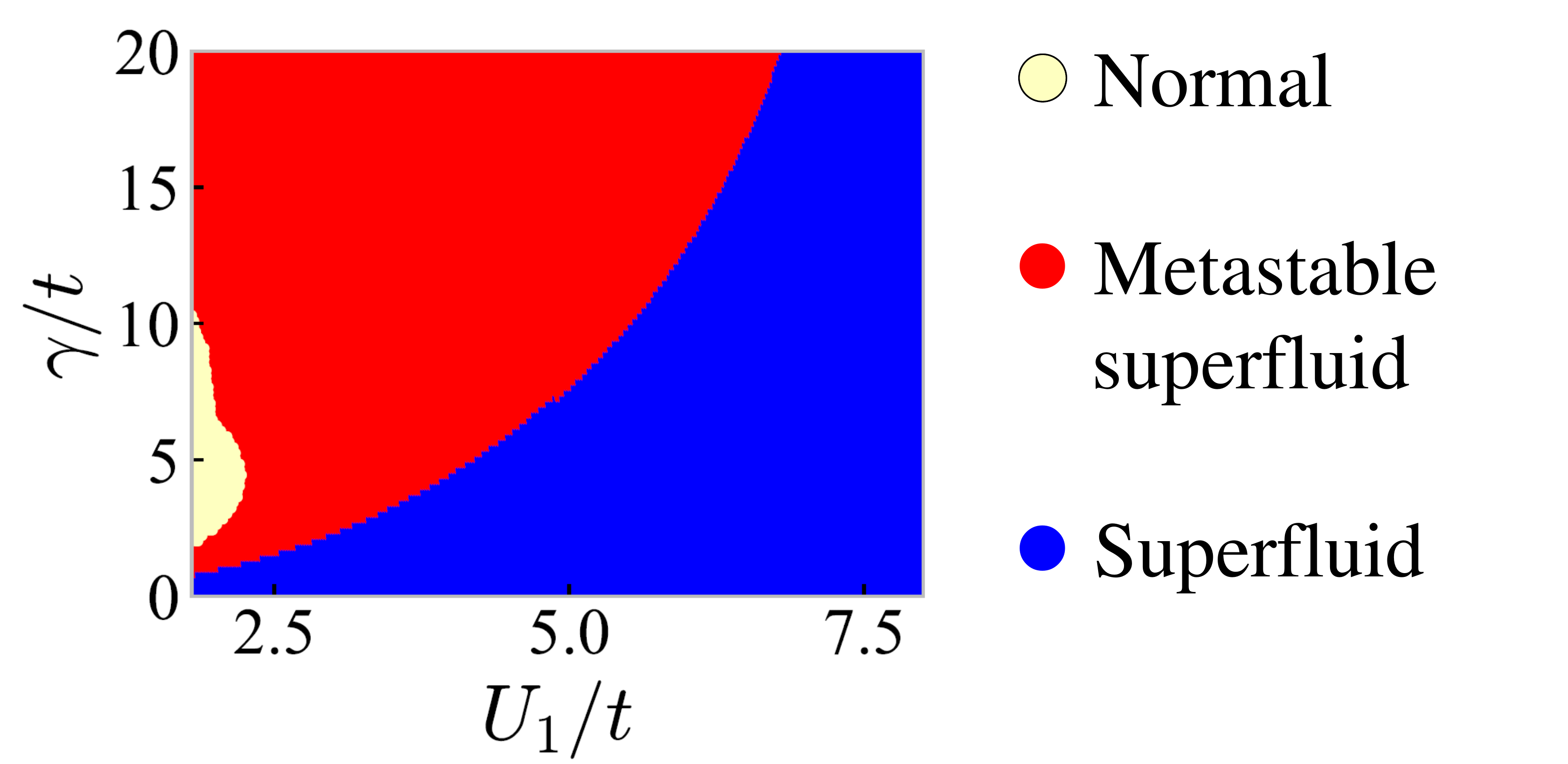}
\caption{Phase diagram of the NH BCS model at $\beta\to\infty$ and $\mu=0$. The yellow region corresponds to the normal state. The red region shows the metastable superfluid state for which a nontrivial solution of the gap equation gives a local energy minimum. The blue region shows the superfluid state corresponding to a nontrivial solution of the gap equation that gives an effective ground state. A region with small $U_1$ is not shown because of the limitation of numerical calculations.}
\label{fig_phasediagramN}
\end{figure}

From these results, we obtain a phase diagram of the NH BCS model as shown in Fig.\ \ref{fig_phasediagramN}. In the blue region, the superfluid state is an effective ground state of the NH BCS Hamiltonian. When the dissipation is increased, the superfluid state remains stable if the attraction is sufficiently strong. When the system enters the red region, the superfluid state becomes metastable with respect to the real part of the energy. The metastable superfluid undergoes an unconventional quantum phase transition due to exceptional points in the case of weak attractions, leading to the disappearance of the superfluid state in the yellow region. Here we remark that a similar phase diagram is obtained from an exact solution of a one-dimensional NH Hubbard model \cite{Nakagawainprep}. Although a similar formulation to obtain the gap equation can be made for a continuum system, the reentrant superfluidity is unique to the lattice system since the localization due to the QZE cannot occur without a lattice. On the other hand, the breakdown of the superfluidity and the related phase transitions can occur in the continuum system.

\textit{Towards experimental realization.}--\
The NH quantum phase transitions can be observed by controlling a two-body loss rate. Since the superfluid is metastable in the red region, it can be realized by slowly increasing dissipation from the blue region. For weak attraction $U_1$, the superfluid undergoes an unconventional phase transition to the normal state due to the exceptional points. To observe the reappearance of the superfluidity under large $\gamma$, we may first prepare a metastable superfluid at large $U_1$ and $\gamma$, and then decrease $U_1$. Finally, the metastablity of the superfluid can be confirmed through comparison of the results between fast and slow increases of the dissipation from $\gamma = 0$. For a detailed discussion including the relevant timescales, see Supplemental Material \cite{Supple}.

These NH superfluids are expected to be realized with ultracold atoms under inelastic collisions. For example, a superfluid of $^{173}$Yb atoms with an orbital Feshbach resonance \cite{Zhang15} offers one such candidate since it is inevitably accompanied by two-body losses as observed experimentally \cite{Hofer15, Pagano15, Cappellini19}. The effect of non-Hermiticity on the superfluid gap can be observed by spectroscopy with Raman transitions between hyperfine levels or a clock transition \cite{Cappellini19, Folling19}. Furthermore, concerning the control of two-body loss rates, introducing dissipation with photoassociation techniques \cite{Tomita17} may also enable the realization of NH superfluids with $^{40}$K and $^{6}$Li \cite{Stewart08, Chin04, Ketterle06, Esslinger06, Bakr18}. Although the strength of dissipation is usually fixed by scattering properties of atoms, dissipation engineering using photoassociation techniques will be a feasible method for realizing the NH fermionic superfluidity.

\textit{Conclusions.}--\
We have investigated how the BCS superfluidity is extended to NH quantum systems under inelastic interactions. We have elucidated some remarkable features unique to the NH fermionic superfluidity, such as exotic Bogoliubov quasiparticles which belong to neither fermions nor bosons and found unconventional quantum phase transitions unique to non-Hermiticity. In particular, for weak attraction, it has been revealed that the superfluidity breaks down with increasing dissipation but shows reentrant behavior as dissipation is further increased. Remarkably, these phase transitions are accompanied by distinctive features of the non-Hermiticity, i.e. the emergence of exceptional points, lines and surfaces in the quasiparticle Hamiltonian for one-, two- and three-dimensions. These characteristic features will play a decisive role in detecting NH phase transitions in experiments. On the other hand, for strong attraction, the superfluid state is not suppressed but enhanced due to the confinement of molecules to single sites via the QZE. While we have focused on a conventional $s$-wave superfluid, $p$-wave, $d$-wave and other exotic superfluids in NH systems will also be relevant for experiments, and merit future investigation.

\begin{acknowledgments}
\textit{Acknowledgments.}--\ 
We are grateful to Yuto Ashida, Shunsuke Furukawa, Tsuneya Yoshida and Yoshiro Takahashi for fruitful discussions. 
This work was supported by KAKENHI (Grants No.\ JP16J05078, No.\ JP16K05501, No.\ JP17J03883, No.\ JP18H01140,  and No.\ JP18H01145) and a Grant-in-Aid for Scientific Research on Innovative Areas (KAKENHI Grant No.\ JP15H05855) from the Japan Society for the Promotion of Science. M.N. was supported by RIKEN Special Postdoctoral Researcher Program. K.A. and K.T. thank JSPS for support from a Research Fellowship for Young Scientists.
\end{acknowledgments}

\bibliography{NHFermiSF.bib}


\clearpage

\renewcommand{\thesection}{S\arabic{section}}
\renewcommand{\theequation}{S\arabic{equation}}
\setcounter{equation}{0}
\renewcommand{\thefigure}{S\arabic{figure}}
\setcounter{figure}{0}

\onecolumngrid
\appendix
\begin{center}
\large{Supplemental Material for}\\
\textbf{"Theory of Non-Hermitian Fermionic Superfluidity with a Complex-Valued Interaction"}
\end{center}


\section{Non-Hermitian mean-field theory of fermionic superfluidity}
The partition function is written in the path-integral representation as
\begin{align}
&Z=\int\mathcal{D}\bar{c}\mathcal{D}c\exp\{-S(\bar{c},c)\},\notag\\
&S(\bar{c},c)=\int_0^{\beta}d\tau\left(\sum_{\bm{k}\sigma}\bar{c}_{\bm{k}\sigma}(\tau)(\partial_{\tau}+\xi_{\bm{k}})c_{\bm{k}\sigma}(\tau)-U\sum_{i}\bar{c}_{i\uparrow}(\tau)\bar{c}_{i\downarrow}(\tau)c_{i\downarrow}(\tau)c_{i\uparrow}(\tau)\right),
\end{align}
where $c$ and $\bar{c}$ are Grassmann variables, and $\tau$ is the imaginary time. After performing the Hubbard-Stratonovich transformation, the partition function is rewritten as
\begin{align}
Z=\int\mathcal{D}\bar{c}\mathcal{D}c\mathcal{D}\bar{\Delta}\mathcal{D}\Delta{e}^{-S(\bar{\Delta},\Delta,\bar{c},c)},
\end{align}
\begin{align}
S(\bar{\Delta},\Delta,\bar{c},c)=\int_0^\beta{d\tau}\left[\sum_{\bm{k}\sigma}\bar{c}_{\bm{k}\sigma}(\tau)(\partial_{\tau}+\xi_{\bm{k}})c_{\bm{k}\sigma}(\tau)+\sum_i\left(\frac{\bar{\Delta}_i(\tau)\Delta_i(\tau)}{U}+\bar{\Delta}_i(\tau)c_{i\downarrow}(\tau)c_{i\uparrow}(\tau)+\Delta_i(\tau)\bar{c}_{i\uparrow}(\tau)\bar{c}_{i\downarrow}(\tau)\right)\right],
\label{eq_actionHS}
\end{align}
where $\bar{\Delta}_i(\tau)$ and $\Delta_i(\tau)$ are auxiliary bosonic fields, which are not necessarily complex conjugate to each other in the saddle-point approximation. Substituting $c_{i\sigma}(\tau)=\frac{1}{\sqrt{\beta{N}}}\sum_{\bm{k},\omega_n}e^{i(\bm{r}_i\cdot\bm{k}-\omega_n\tau)}c_{\bm{k}\sigma}(\omega_n)$, $\bar{c}_{i\sigma}(\tau)=\frac{1}{\sqrt{\beta{N}}}\sum_{\bm{k},\omega_n}e^{-i(\bm{r}_i\cdot\bm{k}-\omega_n\tau)}\bar{c}_{\bm{k}\sigma}(\omega_n)$, $\Delta_i(\tau)=\frac{1}{\sqrt{\beta{N}}}\sum_{\bm{k},\Omega_l}e^{i(\bm{r}_i\cdot\bm{k}-\Omega_l\tau)}\Delta_{\bm{k}}(\Omega_l)$ and $\bar{\Delta}_i(\tau)=\frac{1}{\sqrt{\beta{N}}}\sum_{\bm{k},\Omega_l}e^{-i(\bm{r}_i\cdot\bm{k}-\Omega_l\tau)}\bar{\Delta}_{\bm{k}}(\Omega_l)$ into Eq.~\eqref{eq_actionHS}, we have
\begin{align}
S(\bar{\Delta},\Delta,\bar{c},c)=&\frac{1}{U}\sum_{\bm{k},\Omega_l}\bar{\Delta}_{\bm{k}}(\Omega_l)\Delta_{\bm{k}}(\Omega_l)+\sum_{\bm{k},\omega_n,\sigma}\bar{c}_{\bm{k}\sigma}(\omega_n)(-i\omega_n+\xi_{\bm{k}})c_{\bm{k}\sigma}(\omega_n)\notag\\
&+\sum_{\bm{k},\bm{k}'',\omega_n,\Omega_l}\frac{1}{\sqrt{\beta{N}}}\bar{\Delta}_{\bm{k}''}(\Omega_l)c_{\bm{k}''-\bm{k}\downarrow}(\Omega_l-\omega_n)c_{\bm{k}\uparrow}(\omega_n)\notag\\
&+\sum_{\bm{k},\bm{k}'',\omega_n,\Omega_l}\frac{1}{\sqrt{\beta{N}}}\Delta_{\bm{k}''}(\Omega_l)\bar{c}_{\bm{k}\uparrow}(\omega_n)\bar{c}_{\bm{k}''-\bm{k}\downarrow}(\Omega_l-\omega_n),
\label{eq_actionFT}
\end{align}
where $\omega_n$ and  $\Omega_l$ are the Matsubara frequencies for fermions and bosons, respectively. In the mean-field theory, we ignore the spatial and temporal fluctuations of $\Delta_{\bm{k}''}(\Omega_l)$ and $\bar{\Delta}_{\bm{k}''}(\Omega_l)$ and thus set $\bm{k}''=0$ and $\Omega_l=0$. Then the action is described as 
\begin{align}
S(\bar{c},c,\bar{\Delta},\Delta)=\frac{\beta{N}}{U}\bar{\Delta}\Delta
+\sum_{\bm{k},\omega_n}
\left(
\begin{matrix}
\bar{c}_{\bm{k}\uparrow}(\omega_n)&c_{-\bm{k}\downarrow}(-\omega_n)
\end{matrix}
\right)
\left(
\begin{matrix}
-i\omega_n+\xi_{\bm{k}}&\Delta\\
\bar{\Delta}&-i\omega_n-\xi_{\bm{k}}
\end{matrix}
\right)
\left(
\begin{matrix}
c_{\bm{k}\uparrow}(\omega_n)\\
\bar{c}_{-\bm{k}\downarrow}(-\omega_n)
\end{matrix}
\right),
\label{eq_actionWF}
\end{align}
where $\Delta=\frac{1}{\sqrt{\beta{N}}}\Delta_{\bm{k}=0}(0)$ and $\bar{\Delta}=\frac{1}{\sqrt{\beta{N}}}\bar{\Delta}_{\bm{k}=0}(0)$. Integrating out the fermionic degrees of freedom by using the formula $\int\exp\left[-\sum_{i,j}\bar{c}_iA_{ij}c_j\right]\prod_{i=1}^n\mathcal{D}\bar{c}_i\mathcal{D}c_i=\mathrm{det}A$, we obtain
\begin{align}
Z=\int\mathcal{D}\bar{\Delta}\mathcal{D}{\Delta}e^{-S_{\mathrm{eff}}(\Delta,\bar{\Delta})},
\end{align}
where
\begin{align}
S_\mathrm{eff}(\bar{\Delta},\Delta)
&=\frac{\beta{N}}{U}\bar{\Delta}\Delta-\sum_{\omega_n,\bm{k}}\mathrm{ln}\left\{-\mathrm{det}\left(
\begin{matrix}
-i\omega_n+\xi_{\bm{k}}&\Delta\\
\bar{\Delta}&-i\omega_n-\xi_{\bm{k}}
\end{matrix}
\right)
\right\}\notag\\
&=-\sum_{\omega_n,\bm{k}}\mathrm{ln}(\omega_n^2+\xi_{\bm{k}}^2+\bar{\Delta}{\Delta})+\frac{\beta{N}}{U}\bar{\Delta}{\Delta},
\label{eq_actioneff}
\end{align}
is the effective action. By differentiating
\begin{align}
e^{-S_{\mathrm{eff}}(\Delta,\bar{\Delta})}=\int\mathcal{D}\bar{c}\mathcal{D}c e^{-S(\bar{\Delta},\Delta,\bar{c},c)}
\end{align}
with respect to $\Delta$ and $\bar{\Delta}$ and using the saddle point condition $\partial{S}_\mathrm{eff}/\partial\Delta=\partial{S}_\mathrm{eff}/\partial\bar{\Delta}=0$, the order parameters are given by
\begin{align}
\Delta=-\frac{U}{N}\sum_{\bm{k}}{}_L\langle c_{-\bm{k}\downarrow}c_{\bm{k}\uparrow}\rangle_R,
\label{eq_orderparameterS1}\\
\bar{\Delta}=-\frac{U}{N}\sum_{\bm{k}}{}_L\langle c_{\bm{k}\uparrow}^\dagger{c}_{-\bm{k}\downarrow}^\dagger\rangle_R.
\label{eq_orderparameterS2}
\end{align}
Here, we have used the expectation values defined as
\begin{align}
&\sum_{\bm{k}}{}_L\langle c_{-\bm{k}\downarrow}c_{\bm{k}\uparrow}\rangle_R=\frac{\int\mathcal{D}\bar{c}\mathcal{D}c\sum_{\bm{k},\omega_n}c_{-\bm{k}\downarrow}(-\omega_n)c_{\bm{k}\uparrow}(\omega_n)e^{-S}}{\int\mathcal{D}\bar{c}\mathcal{D}ce^{-S}},\notag\\
&\sum_{\bm{k}}{}_L\langle c_{\bm{k}\uparrow}^\dagger{c}_{-\bm{k}\downarrow}^\dagger\rangle_R=\frac{\int\mathcal{D}\bar{c}\mathcal{D}c\sum_{\bm{k},\omega_n}\bar{c}_{\bm{k}\uparrow}(\omega_n)\bar{c}_{-\bm{k}\downarrow}(-\omega_n)e^{-S}}{\int\mathcal{D}\bar{c}\mathcal{D}ce^{-S}}.
\end{align}
By using Eqs. \eqref{eq_orderparameterS1} and \eqref{eq_orderparameterS2}, we decouple the NH Hamiltonian \eqref{eq_Hubbard_k} and obtain
\begin{align}
H_{\mathrm{MF}}
=\sum_{\bm{k}}
\left(
\begin{matrix}
c_{\bm{k}\uparrow}^\dagger&c_{-\bm{k}\downarrow}
\end{matrix}
\right)
\left(
\begin{matrix}
\xi_{\bm{k}}&\Delta\\
\bar{\Delta}&-\xi_{\bm{k}}
\end{matrix}
\right)
\left(
\begin{matrix}
c_{\bm{k}\uparrow}\\
c_{-\bm{k}\downarrow}^\dagger
\end{matrix}
\right)
=\sum_{\bm{k}}
\left(
\begin{matrix}
\bar{\gamma}_{\bm{k}\uparrow}&\gamma_{-\bm{k}\downarrow}
\end{matrix}
\right)
\left(
\begin{matrix}
E_{\bm{k}}&0\\
0&-E_{\bm{k}}
\end{matrix}
\right)
\left(
\begin{matrix}
\gamma_{\bm{k}\uparrow}\\
\bar{\gamma}_{-\bm{k}\downarrow}
\end{matrix}
\right),
\end{align}
where the quasiparticle operators are given by
\begin{align}
&\bar{\gamma}_{\bm{k}\uparrow}=u_{\bm{k}}c_{\bm{k}\uparrow}^\dagger-\bar{v}_{\bm{k}}{c}_{-\bm{k}\downarrow},\notag\\
&\bar{\gamma}_{-\bm{k}\downarrow}=\bar{v}_{\bm{k}}c_{\bm{k}\uparrow}+u_{\bm{k}}c_{-\bm{k}\downarrow}^\dagger,\notag\\
&\gamma_{\bm{k}\uparrow}=u_{\bm{k}}c_{\bm{k}\uparrow}-v_{\bm{k}}c_{-\bm{k}\downarrow}^\dagger,\notag\\
&\gamma_{-\bm{k}\downarrow}=v_{\bm{k}}c_{\bm{k}\uparrow}^\dagger+u_{\bm{k}}{c}_{-\bm{k}\downarrow},
\end{align}
and the coefficients are given by
\begin{align}
u_{\bm{k}}=\sqrt{\frac{E_{\bm{k}}+\xi_{\bm{k}}}{2E_{\bm{k}}}},\hspace{0.2cm}
v_{\bm{k}}=-\sqrt{\frac{(E_{\bm{k}}-\xi_{\bm{k}})}{2E_{\bm{k}}}}\frac{\sqrt{\Delta}}{\sqrt{\bar{\Delta}}},
\hspace{0.2cm}
\bar{v}_{\bm{k}}=-\sqrt{\frac{(E_{\bm{k}}-\xi_{\bm{k}})}{2E_{\bm{k}}}}\frac{\sqrt{\bar{\Delta}}}{\sqrt{\Delta}}.
\end{align}


\section{Chemical potential in non-Hermitian systems}
Here we clarify how the chemical potential is determined in non-Hermitian quantum systems. A density matrix $\rho$ of an atomic gas is decomposed into sectors, each of which has a definite particle number $M$ as
\begin{equation}
\rho=\sum_M \rho^{(M)}.
\end{equation}
In the short-time dynamics of the atomic gas, the time evolution of the density matrix $\rho^{(M)}$ is described by the NH Hamiltonian $H_{\mathrm{eff}}$ restricted to the $M$-particle Hilbert space \cite{Durr09}. An effective energy spectrum in the $M$-particle Hilbert space can be extracted by tuning the chemical potential so that the mean particle number in the non-Hermitian ensemble \eqref{eq_partitionfunction} in the main text is
\begin{equation}
M=\frac{1}{Z}\sum_n ~_L\langle E_n|\hat{N}e^{-\beta H_{\mathrm{eff}}}|E_n\rangle_R,
\label{eq_ensemble}
\end{equation}
where $\hat{N}=\sum_{\bm{k},\sigma} c_{\bm{k}\sigma}^\dag c_{\bm{k}\sigma}$ is the particle-number operator. In the $\beta\to\infty$ limit, Eq.~\eqref{eq_ensemble} is calculated as
\begin{equation}
M=~_L\langle\mathrm{BCS}|\hat{N}|\mathrm{BCS}\rangle_R=\sum_{\bm{k}}\left(1-\frac{\xi_{\bm{k}}}{E_{\bm{k}}}\right),
\label{eq_ensemble2}
\end{equation}
where the BCS ground states \eqref{eq_BCSR} and \eqref{eq_BCSL} in the main text are used. For a cubic or square lattice considered in the main text, a half-filling condition $M=N$ is achieved by setting $\mu=0$ in Eq.~\eqref{eq_ensemble2}.

We note that the expectation value of particle numbers taken by the right BCS eigenstates does not coincide with Eq.~\eqref{eq_ensemble2}:
\begin{align}
\frac{~_R\langle\mathrm{BCS}|\hat{N}|\mathrm{BCS}\rangle_R}{~_R\langle\mathrm{BCS}|\mathrm{BCS}\rangle_R}&=\sum_{\bm{k}}\frac{2|v_{\bm{k}}|^2}{|u_{\bm{k}}|^2+|v_{\bm{k}}|^2}\notag\\
&=2\sum_{\bm{k}}\frac{\left|1-\frac{\xi_{\bm{k}}}{E_{\bm{k}}}\right|}{\left|1+\frac{\xi_{\bm{k}}}{E_{\bm{k}}}\right|+\left|1-\frac{\xi_{\bm{k}}}{E_{\bm{k}}}\right|},
\end{align}
since the BCS states do not conserve the particle number. However, this is not a contradiction since in the $M$-particle Hilbert space, the number-conserving states
\begin{align}
|\Psi_M\rangle_R&\equiv\frac{1}{\mathcal{N}}\left(\sum_{\bm{k}}\alpha_{\bm{k}}c^\dag_{\bm{k}\uparrow} c^\dag_{-\bm{k}\downarrow}\right)^{M/2}\ket{0}\notag\\
&\propto\int_0^{2\pi}d\theta e^{-iM\theta/2}\prod_{\bm{k}}(u_{\bm{k}}+v_{\bm{k}}(\theta)c^\dag_{\bm{k}\uparrow} c^\dag_{-\bm{k}\downarrow})\ket{0},\\
|\Psi_M\rangle_L&\equiv\frac{1}{\mathcal{N}}\left(\sum_{\bm{k}}\alpha_{\bm{k}}^*c^\dag_{\bm{k}\uparrow} c^\dag_{-\bm{k}\downarrow}\right)^{M/2}\ket{0}\notag\\
&\propto\int_0^{2\pi}d\theta e^{-iM\theta/2}\prod_{\bm{k}}(u_{\bm{k}}^*+\bar{v}_{\bm{k}}^{*}(\theta)c^\dag_{\bm{k}\uparrow} c^\dag_{-\bm{k}\downarrow})\ket{0}, 
\end{align}
are realized instead of the BCS states \cite{Leggett91, Leggett06}. Here, $\alpha_{\bm{k}}\equiv u_{\bm{k}}/v_{\bm{k}}(\theta=0)$. For these states, the particle number does not depend on the choice of an expectation value:
\begin{equation}
\frac{~_L\langle\Psi_M|\hat{N}|\Psi_M\rangle_R}{~_L\langle\Psi_M|\Psi_M\rangle_R}=\frac{~_R\langle\Psi_M|\hat{N}|\Psi_M\rangle_R}{~_R\langle\Psi_M|\Psi_M\rangle_R}=M.
\end{equation}


\section{Calculation of the condensation energy of the superfluid state}
We here discuss how to obtain the condensation energy of the superfluid state for consideration of its stability. Using Eq.\ \eqref{eq_actioneff}, the condensation energy is given by the difference in energy between the superfluid and normal states as
\begin{align}
E
&=\frac{1}{\beta}\left(S_\mathrm{eff}(\Delta_0e^{i\theta},\Delta_0e^{-i\theta})-S_\mathrm{eff}(0,0)\right)\notag\\
&=\frac{N}{U}\Delta_0^2+\frac{1}{\beta}\sum_{\omega_n\bm{k}}\left\{\log(i\omega_n+|\xi_{\bm{k}}|)+\log(-i\omega_n+|\xi_{\bm{k}}|)-\log(i\omega_n+\sqrt{\xi_{\bm{k}}^2+\Delta_0^2})-\log(-i\omega_n+\sqrt{\xi_{\bm{k}}^2+\Delta_0^2})\right\}\notag\\
&=\frac{N}{U}\Delta_0^2-\frac{1}{\beta}\sum_{\omega_n,\bm{k}}\log\left(1+\frac{\Delta_0^2}{\omega_n^2+\xi_{\bm{k}}^2}\right).
\label{eq_CE}
\end{align}
\begin{figure}[t]
\includegraphics[width=8.5cm]{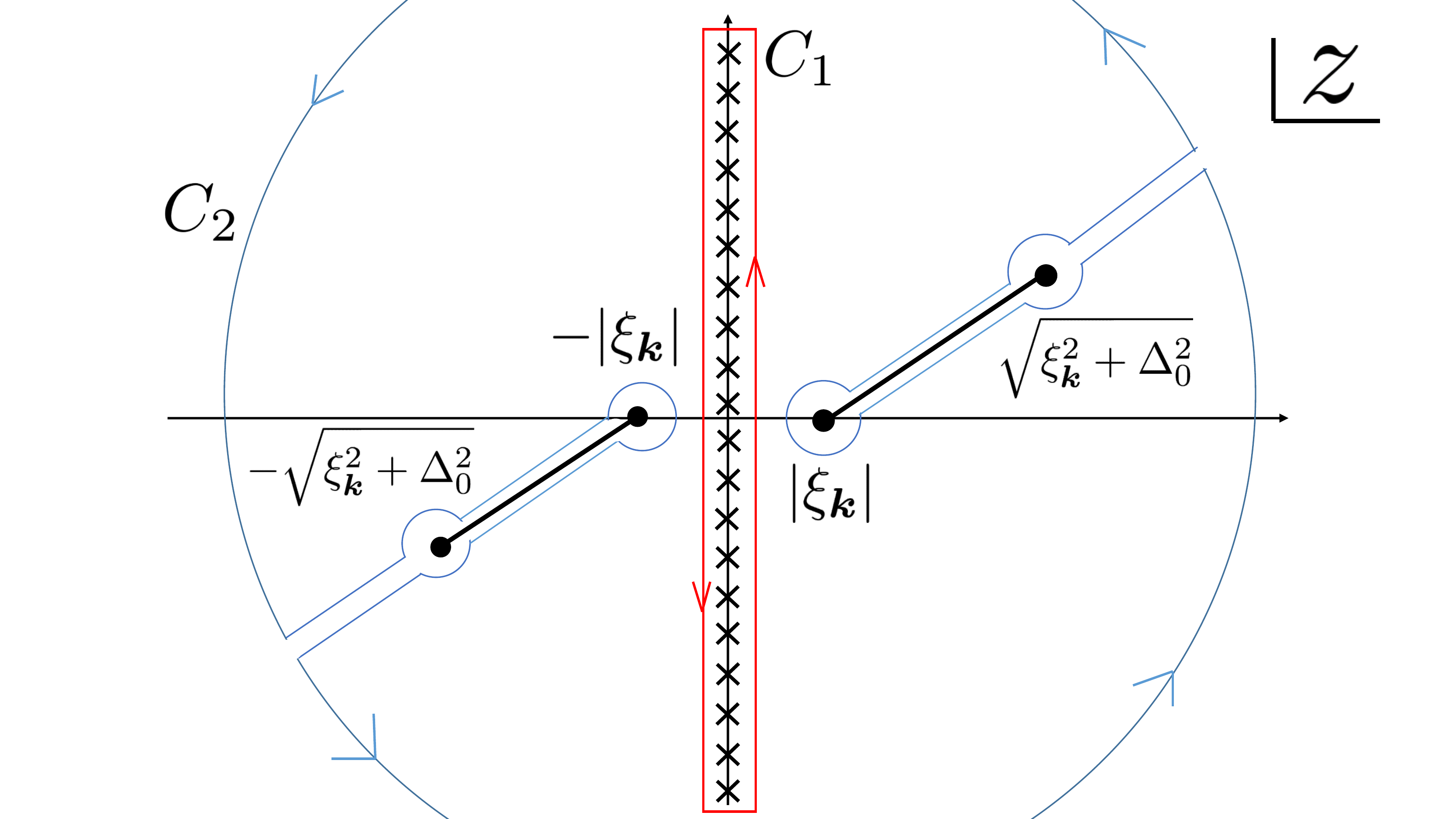}
\caption{Integration contour in the calculation of the condensation energy.}
\label{fig_integration}
\end{figure}
Using the integration contour in Fig.\ \ref{fig_integration}, we can calculate the sum over the Matsubara frequency as
\begin{align}
E=\frac{N}{U}\Delta_0^2+\frac{1}{2\pi{i}}\sum_{\bm{k}}\oint_{C_2}dz\log\left(1+\frac{\Delta_0^2}{-z^2+\xi_{\bm{k}}^2}\right)f(z),
\label{eq_energyintegrand}
\end{align}
where $f(z)=(e^{\beta{z}}+1)^{-1}$ is the Fermi distribution function.
We here note that the integrand in Eq.~\eqref{eq_energyintegrand} has a branch cut on the lines connecting the branch points of Eq.\ \eqref{eq_CE} as shown in Fig.\ \ref{fig_integration}. We thus obtain
\begin{align}
E=\frac{N}{U}\Delta_0^2-\sum_{\bm{k}}\int_{|\xi_{\bm{k}}|}^{\sqrt{\xi_{\bm{k}}^2+\Delta_0^2}}dz\tanh\frac{\beta{z}}{2}.
\end{align}
In the $\beta\to\infty$ limit, the condensation energy is given by
\begin{align}
E=\frac{N}{U}\Delta_0^2-\sum_{\bm{k}}(\sqrt{\xi_{\bm{k}}^2+\Delta_0^2}-|\xi_{\bm{k}}|).
\label{eq_condenergy}
\end{align}

\begin{figure}[h]
\centering
\begin{minipage}[c]{0.37\textwidth}
\centering
\includegraphics[width=6.5cm]{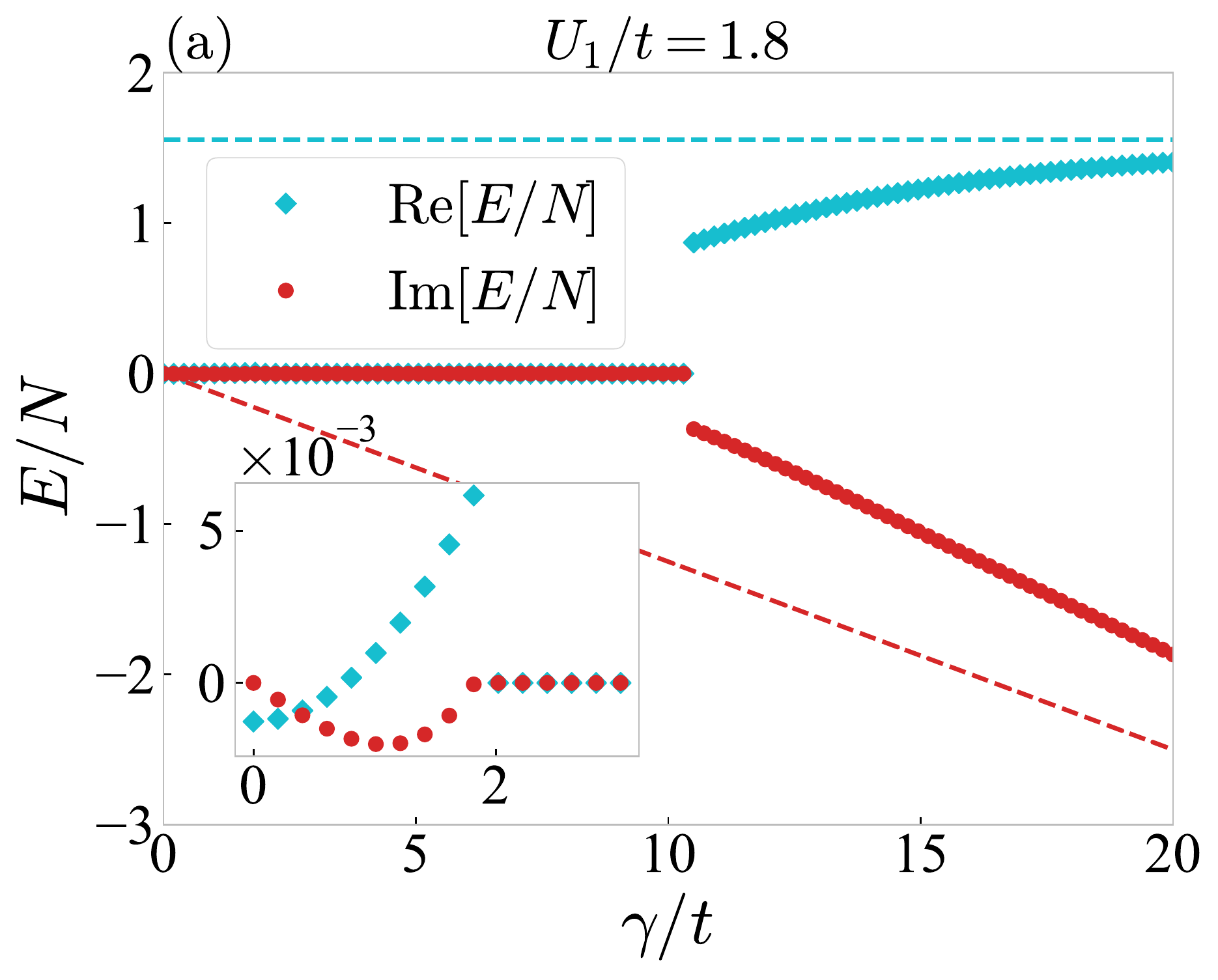}
\end{minipage}
\begin{minipage}[c]{0.37\textwidth}
\centering
\includegraphics[width=6.5cm]{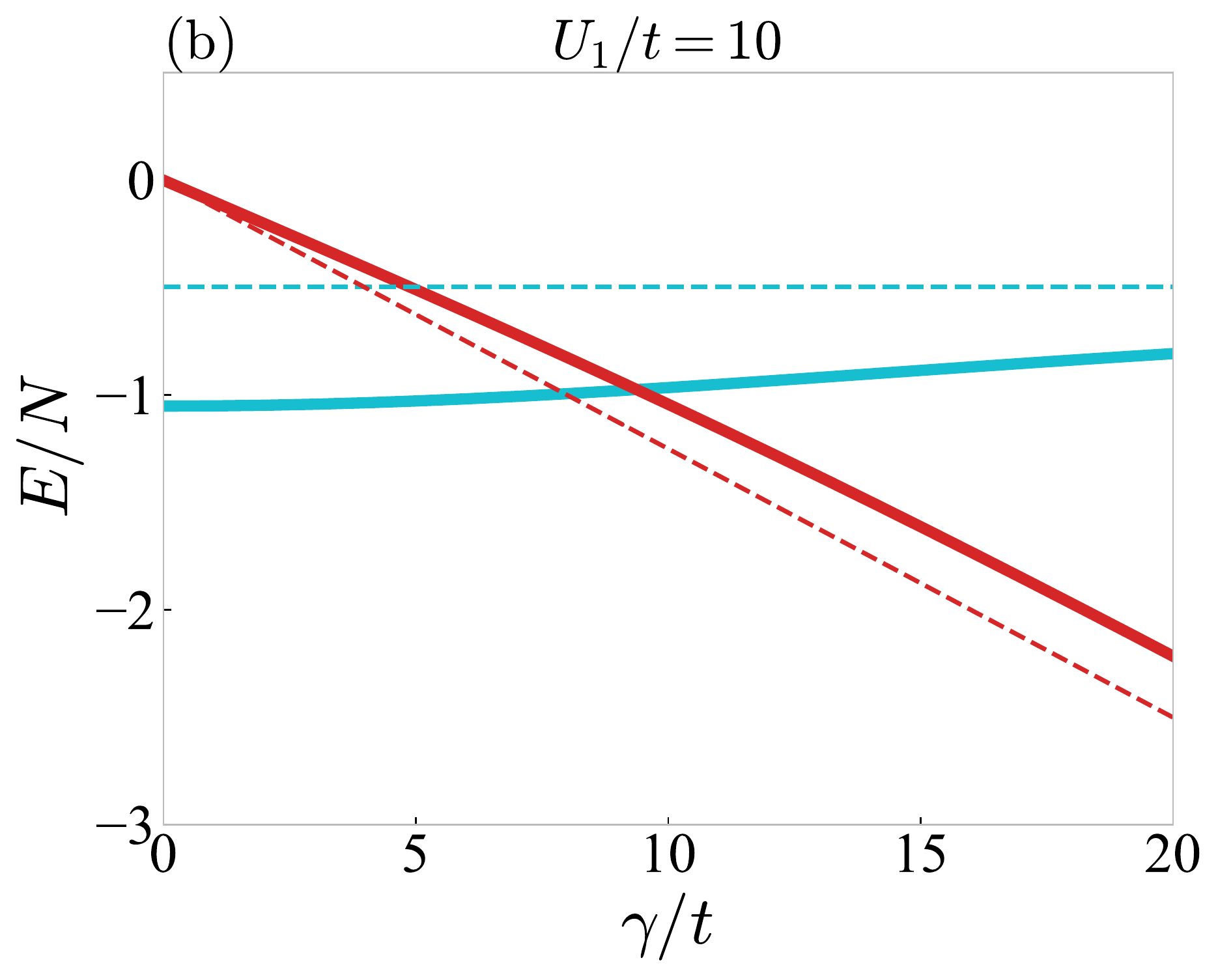}
\end{minipage}
\caption{Condensation energy \eqref{eq_condenergy} as a function of $\gamma/t$.  We here assume the cubic lattice, and the interaction and the chemical potential are set to the same values as in Fig.\ \ref{fig_gapN} in the main text, respectively. The dashed lines show the asymptotic behavior in the strong-dissipation limit. The inset in (a) shows an enlarged view of the weak-dissipation regime.}
\label{fig_condenergy}
\end{figure}

Figure \ref{fig_condenergy} shows the condensation energy for the parameters corresponding to those in Fig.\ \ref{fig_gapN} in the main text. For $U_1/t=1.8$, we find that the superfluid state is energetically stable only in the weak-dissipation region. For a region where the real part of the condensation energy is positive, the nontrivial solution of the NH gap equation gives a local minimum of the real part of energy, leading to a metastable superfluid solution. We also find that the nontrivial solution of the gap equation in the strong-dissipation regime gives a metastable state because the energy is higher than that of the trivial solution. On the other hand, for $U_1/t=10$, the superfluid is always energetically stable, and we can observe the steady enhancement of the superfluid gap.

\section{Numerical results in an intermediate regime $U_1/t=2.5$}
\begin{figure}[h]
\centering
\begin{minipage}[c]{0.37\textwidth}
\centering
\includegraphics[width=6.5cm]{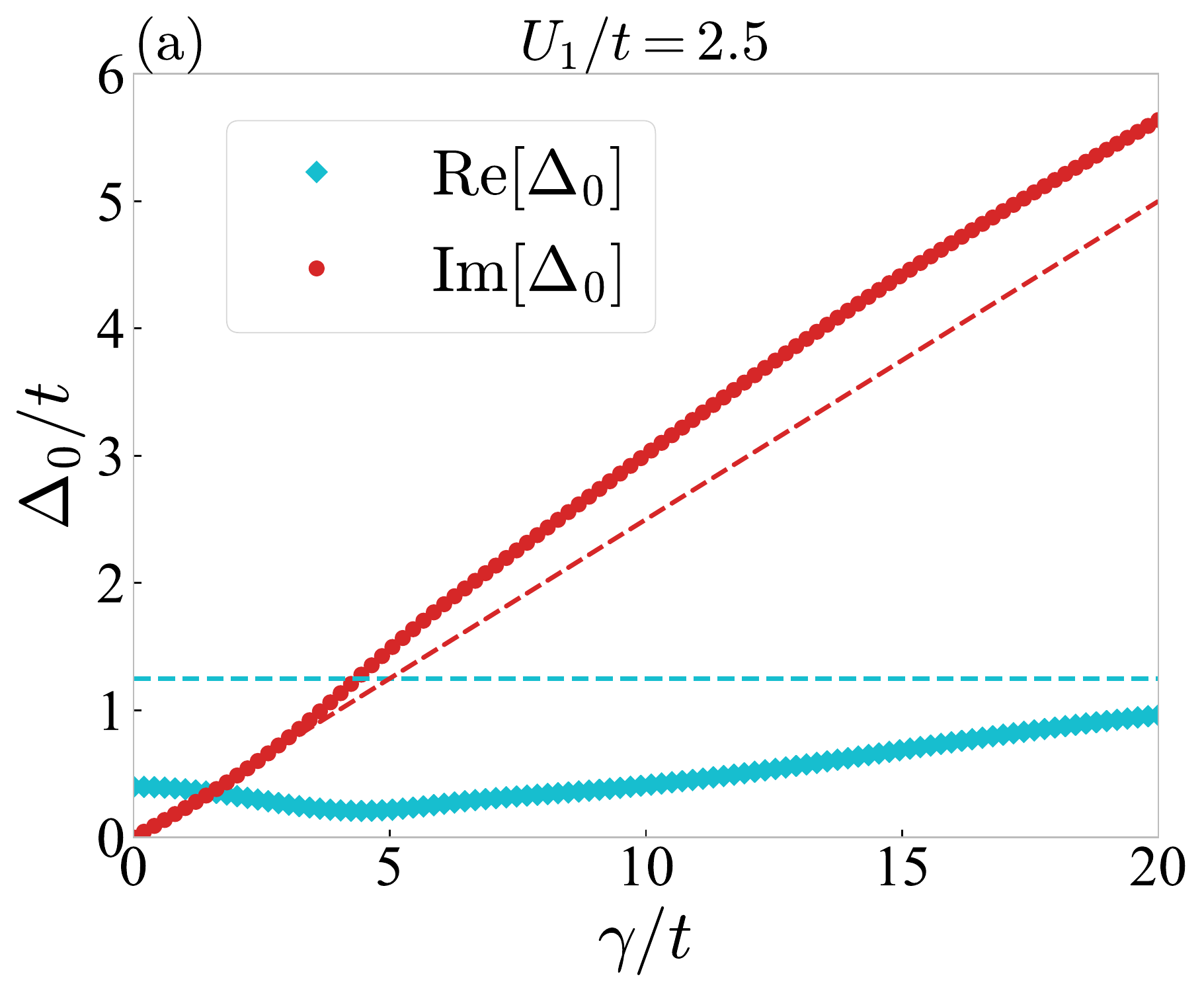}
\end{minipage}
\begin{minipage}[c]{0.37\textwidth}
\centering
\includegraphics[width=6.5cm]{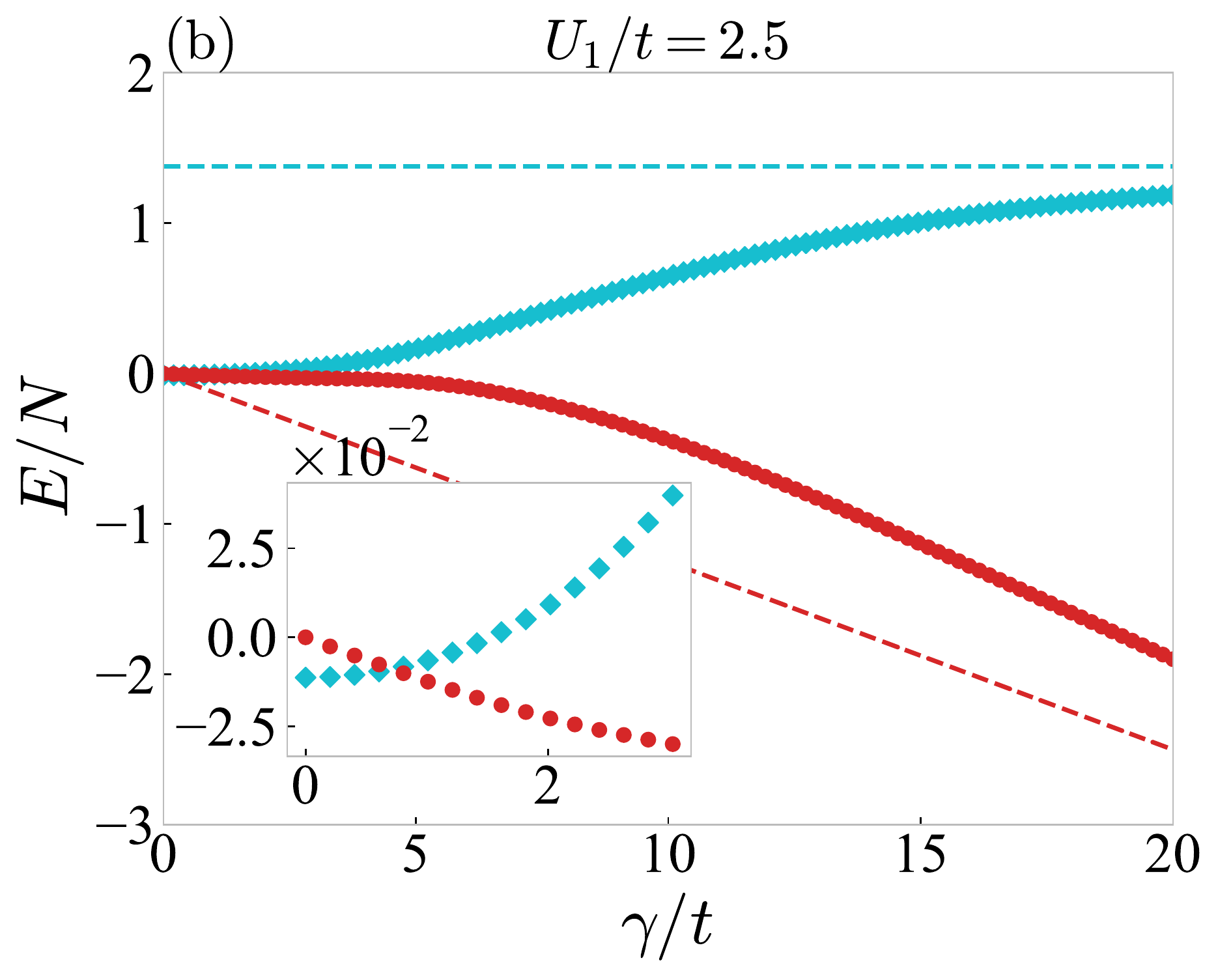}
\end{minipage}
\caption{(a) Superfluid gap obtained from the NH gap equation at $\beta\to\infty$ and (b) the condensation energy \eqref{eq_condenergy} for $U_1=2.5$. The chemical potential measured from the Fermi energy is set to zero and a cubic lattice is assumed as in Fig.\ \ref{fig_gapN}. The dashed lines show the asymptotic behavior in the strong-dissipation limit. The inset in (b) shows an enlarged view in the weak-dissipation regime.}
\label{fig_GE2p5}
\end{figure}

We have conducted numerical calculations for an intermediate strength of the interaction $U_1/t=2.5$ as shown in Fig.\ \ref{fig_GE2p5}. We see from Fig.\ \ref{fig_GE2p5}(a) that the real part of the superfluid gap is suppressed by dissipation but does not vanish. Moreover, it is even enhanced and eventually exceeds the value of the Hermitian limit as the dissipation increases, which is attributed to the confinement of Cooper pairs to individual sites due to the QZE. On the other hand, from Fig.\ \ref{fig_GE2p5}(b), the superfluid state under sufficiently strong dissipation becomes metastable due to the positive condensation energy. While the superfluid does not exhibit an unconventional phase transition with exceptional points in this case, a remnant of the phase transition in a weak attraction can be seen as the minimum of $\mathrm{Re}\Delta_0$ in Fig.\ \ref{fig_GE2p5}(a). 
\begin{figure}[b]
\includegraphics[width=10.5cm]{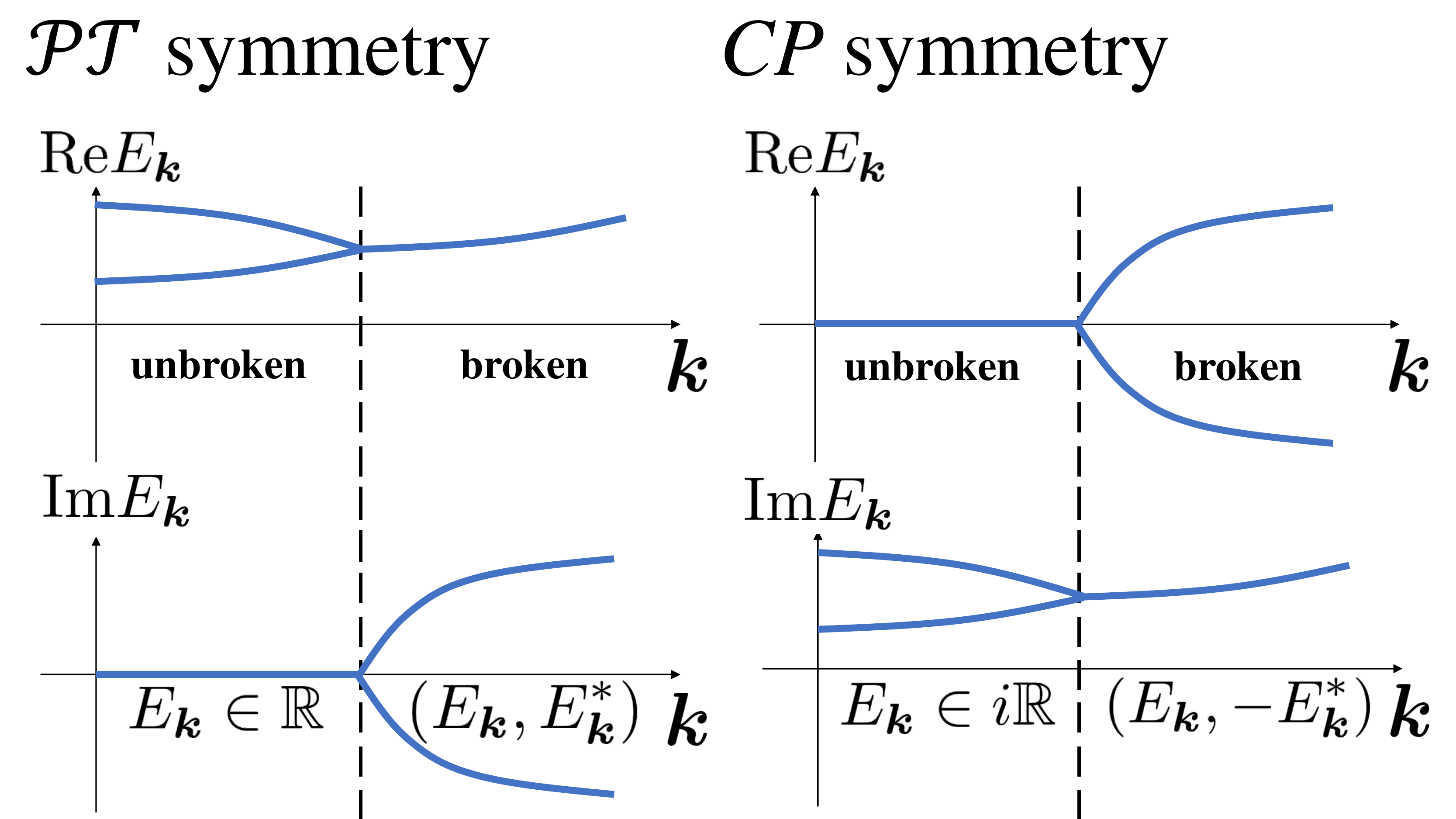}
\caption{Schematic diagrams for the eigenspectrum of a Hamiltonian that has $\mathcal{PT}$ symmetry (left) or $CP$ symmetry (right). The dashed lines indicate the points of spontaneous symmetry breaking.}
\label{fig_symmetry}
\end{figure}
\section{Spontaneous $CP$ symmetry breaking as an origin of exceptional points}
At the breakdown and restoration points of the superfluid gap, the gap is pure imaginary and the mean-field Hamiltonian $H_{\mathrm{MF}}(\bm{k})=\epsilon_{\bm{k}}\sigma_z+i\mathrm{Im}\Delta_0\sigma_x$ satisfies the following $CP$ symmetry \cite{Okugawa19}
\begin{align}
CPH_{\mathrm{MF}}(\bm{k})(CP)^{-1}=-H_{\mathrm{MF}}(\bm{k}),
\end{align}
where the operator $CP$ is given by $CP=\sigma_xK$. This is similar to parity-time ($\mathcal{PT}$) symmetry \cite{Bender98, Bender07}, which is expressed for a general Hamiltonian $H(\bm{k})$ as
\begin{align}
\mathcal{PT}H(\bm{k})(\mathcal{PT})^{-1}=H(\bm{k}),
\end{align}
and they are indeed equivalent to each other as can be seen if we multiply the Hamiltonian by $i$ \cite{Kawabata19}. Consequently, as the $\mathcal{PT}$ symmetry dictates that the eigenspectrum of the Hamiltonian appears as real or complex-conjugate pairs \cite{Bender07}, the $CP$ symmetry dictates that the eigenspectrum appears as pure imaginary or anti-complex-conjugate pairs (Fig.\ \ref{fig_symmetry}). One can easily confirm that the $CP$ symmetry is unbroken if the eigenspectrum is pure imaginary; otherwise it is spontaneously broken.

A generic two-band Hamiltonian with the $CP$ symmetry can be written as
\begin{align}
H(\bm{k})=ia(\bm{k})\sigma_0 + b_z(\bm{k})\sigma_z + id_x(\bm{k})\sigma_x + id_y(\bm{k})\sigma_y,
\end{align}
where $a(\bm{k}), b_z(\bm{k}), d_x(\bm{k}), d_y(\bm{k})\in\mathbb{R}$ ($\sigma_0$ is the $2\times2$ identity matrix). Since its eigenvalues are given by $ia(\bm{k})\pm\sqrt{b_z(\bm{k})^2-d_x(\bm{k})^2-d_y(\bm{k})^2}$, the spontaneous $CP$-symmetry-breaking transition occurs when
\begin{align}
b_z(\bm{k})^2-d_x(\bm{k})^2-d_y(\bm{k})^2 = 0,
\label{eq_EPcondition}
\end{align}
which is accompanied by the emergence of exceptional points. We note 
that only a single condition \eqref{eq_EPcondition} is needed for the exceptional point, while two conditions are needed for an exceptional point in a system without any symmetry \cite{Yoshida19, Budich19, Okugawa19, Zhou19}. Because of this symmetry constraint, the exceptional points form a $(d-1)$-dimensional surface in $d$-dimensional systems. In our case, we have $a(\bm{k})=d_y(\bm{k})=0, b_z(\bm{k})=\xi_{\bm{k}}$, and $d_x(\bm{k})=\mathrm{Im}\Delta_0$ from the mean-field Hamiltonian $H_\mathrm{MF}(\bm{k})$. The quasiparticle energy spectra and exceptional points depicted in Fig.~\ref{fig_exceptional} in the main text are consistent with the above general argument.
\begin{figure}[b]
\includegraphics[width=10.5cm]{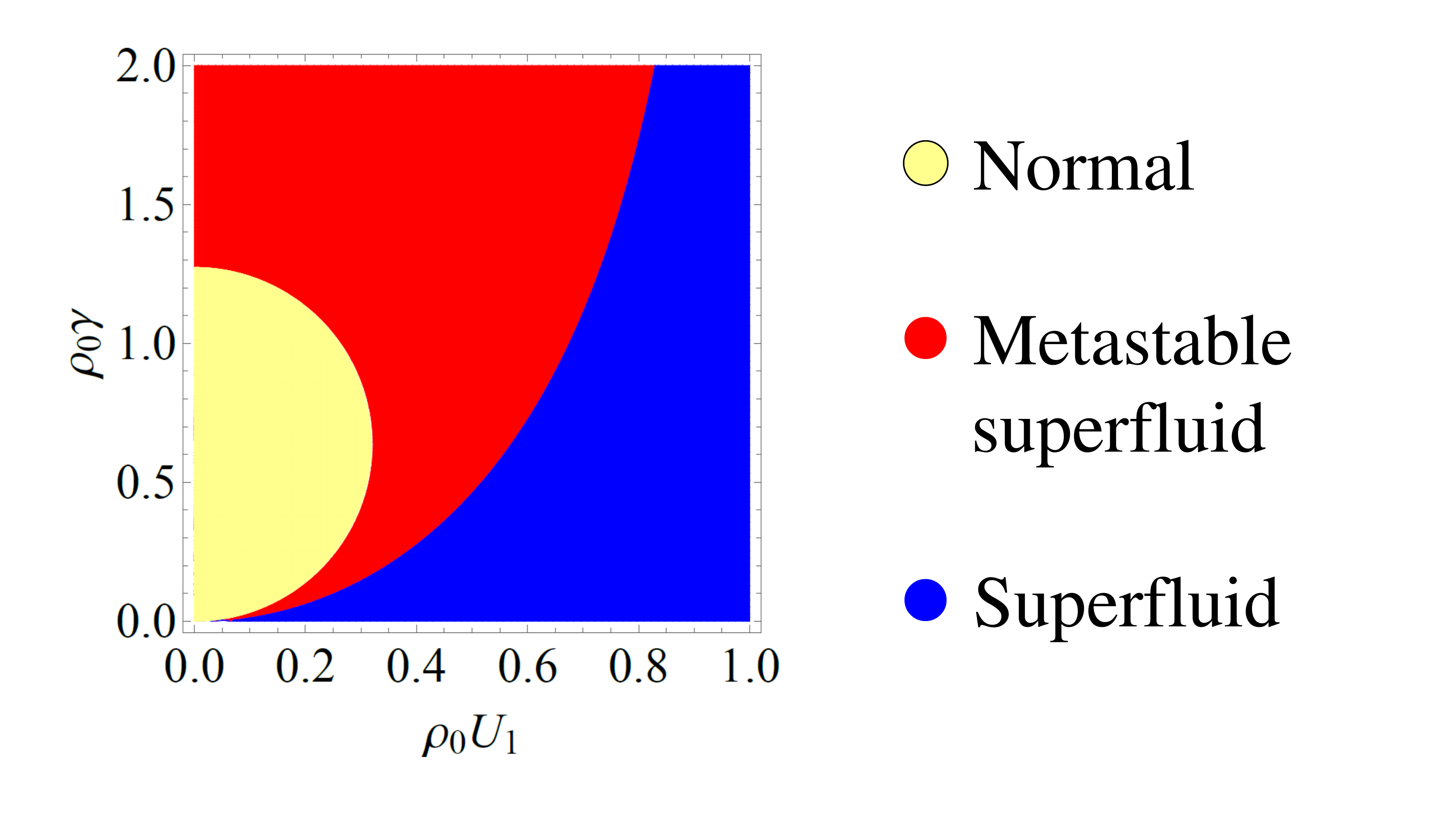}
\caption{Phase diagram obtained from Eqs.\ \eqref{eq_gapA} and \eqref{eq_energyA}. The yellow region corresponds to the normal state where the gap equation has only a trivial solution $\Delta_0=0$. The red region indicates a metastable superfluid state which corresponds to nontrivial solutions of the gap equation but is not an effective ground state. The blue region corresponds to a superfluid phase where a nontrivial solution of the gap equation gives an effective ground state.}
\label{fig_phaseA}
\end{figure}

\section{Analytical calculation of the gap equation and the condensation energy at a constant density of states}
The gap equation (Eq.\ \eqref{eq_gap} in the main text) can be solved analytically if the density of states $\rho_0$ is constant. Under this assumption, the gap equation in the $\beta\to\infty$ limit reads
\begin{align}
\log\left|\frac{\sqrt{\omega_D^2+\Delta_0^2}+\omega_D}{\Delta_0}\right|+i\mathrm{Arg}\frac{\sqrt{\omega_D^2+\Delta_0^2}+\omega_D}{\Delta_0}
=\frac{1}{\rho_0}\left(\frac{U_1}{|U|^2}-i\frac{\gamma}{2|U|^2}\right),
\end{align}
where $\omega_D=1/2\rho_0$ is the energy cutoff. Here, we set a branch cut of $\sqrt{z}$ and $\log{z}$ to $z\in(-\infty,0)$ and assume that $\mathrm{Re}\Delta_0$ is positive without loss of generality. Results obtained below do not depend on the choice of the branch cut. Then, the argument is restricted to $\mathrm{Arg}\frac{\sqrt{\omega_D^2+\Delta_0^2}+\omega_D}{\Delta_0}\in(-\frac{\pi}{2},\frac{\pi}{2})$ and the gap equation has nontrivial solutions if and only if $U_1$ satisfies  $\rho_0U_1<\frac{1}{\pi}$ and $\gamma$ satisfies $0\le\gamma\le\gamma^{c_1}$ or $\gamma\ge\gamma^{c_2}$, where $\gamma^{c_1}$ and $\gamma^{c_2}$ are determined from $(\rho_0\pi{U}_1)^2+(\rho_0\pi\gamma/2-1)^2=1$ as
\begin{align}
&\gamma^{c_1}=\frac{2}{\rho_0\pi}(1-\sqrt{1-(\rho_0\pi{U}_1)^2}),\\
&\gamma^{c_2}=\frac{2}{\rho_0\pi}(1+\sqrt{1-(\rho_0\pi{U}_1)^2}).
\label{eq_critical}
\end{align}
Under the above conditions, the nontrivial solution of the gap equation at $\beta\to\infty$ is given by
\begin{align}
\Delta_0=\frac{\omega_D}{\sinh\left(\frac{1}{\rho_0U}\right)}.
\label{eq_gapA}
\end{align}
This gives the behavior consistent with Fig.\ \ref{fig_gapN} in the main text.

Next, we calculate the condensation energy \eqref{eq_condenergy} for a constant density of states. In this case, Eq.~\eqref{eq_condenergy} can be rewritten as
\begin{align}
\frac{E}{N}=\frac{\Delta_0^2}{U}-\frac{1}{2}\rho_0\omega_D^2\left(-2+2\sqrt{1+\alpha^2}-\alpha^2\left(\mathrm{Log}\alpha^2-2\mathrm{Log}(1+\sqrt{1+\alpha^2})\right)\right),
\label{eq_energyA}
\end{align}
where $\alpha=\frac{\Delta_0}{\omega_D}$. This gives the behavior consistent with Fig.\ \ref{fig_condenergy}.

Finally, we show a phase diagram in Fig.\ \ref{fig_phaseA} obtained from the analytical solutions of the gap equation with a constant density of states. In Fig.\ \ref{fig_phaseA}, the yellow region corresponds to the normal state, where the gap equation has only a trivial solution $\Delta_0=0$. This phase is surrounded by the red region, where the gap equation has the nontrivial solution \eqref{eq_gapA} which corresponds to the metastable superfluid state, since the nontrivial solution gives a local minimum of the real part of energy due to a positive condensation energy. When the attractive interaction $U_1$ is sufficiently strong, the system is in the superfluid phase (blue region in Fig.\ \ref{fig_phaseA}), where the nontrivial solution of the gap equation gives an effective ground state. The phase diagram is qualitatively consistent with Fig.\ \ref{fig_phasediagramN} in the main text.

\section{Details of experimental setups for probing the non-Hermitian superfluid}
Here we discuss an experimental setup for probing the NH superfluidity. As mentioned in the main text, the NH dynamics is realized when we neglect the quantum-jump term in the master equation \eqref{eq_master}. To fulfill this condition, the loss rate $\gamma$ should be much smaller than the energy scale that governs the timescale for relaxation towards a quasi-equilibrium state. In superfluid states, the thermalization proceeds in a timescale of hopping of atoms to neighboring sites \cite{Ketterle06}. Thus, for a BCS superfluid in which $\gamma, U_1 \ll t$ is satisfied, the NH breakdown of superfluid may be observed. We note that, as inferred from Figs.~\ref{fig_phasediagramN} and \ref{fig_phaseA}, the breakdown of superfluid due to exceptional points can be induced by small dissipation in the BCS regime, which justifies the NH dynamics. On the other hand, for the reentrant superfluidity, which is another unique phenomenon in the NH superfluid, dissipation larger than the hopping is required (see Figs.~\ref{fig_phasediagramN} and \ref{fig_phaseA}). This seems to invalidate the assumption for the ignorance of the quantum-jump term. However, as mentioned in the main text, the physics behind the reentrant superfluidity is the QZE, which suppresses the hopping and facilitates the formation of on-site molecules. The manifestation of the QZE can be observed by using the following protocol. Hence, a large dissipation is not incompatible with the ignorance of the quantum-jump term.

\begin{figure}[h]
\includegraphics[width=10.5cm]{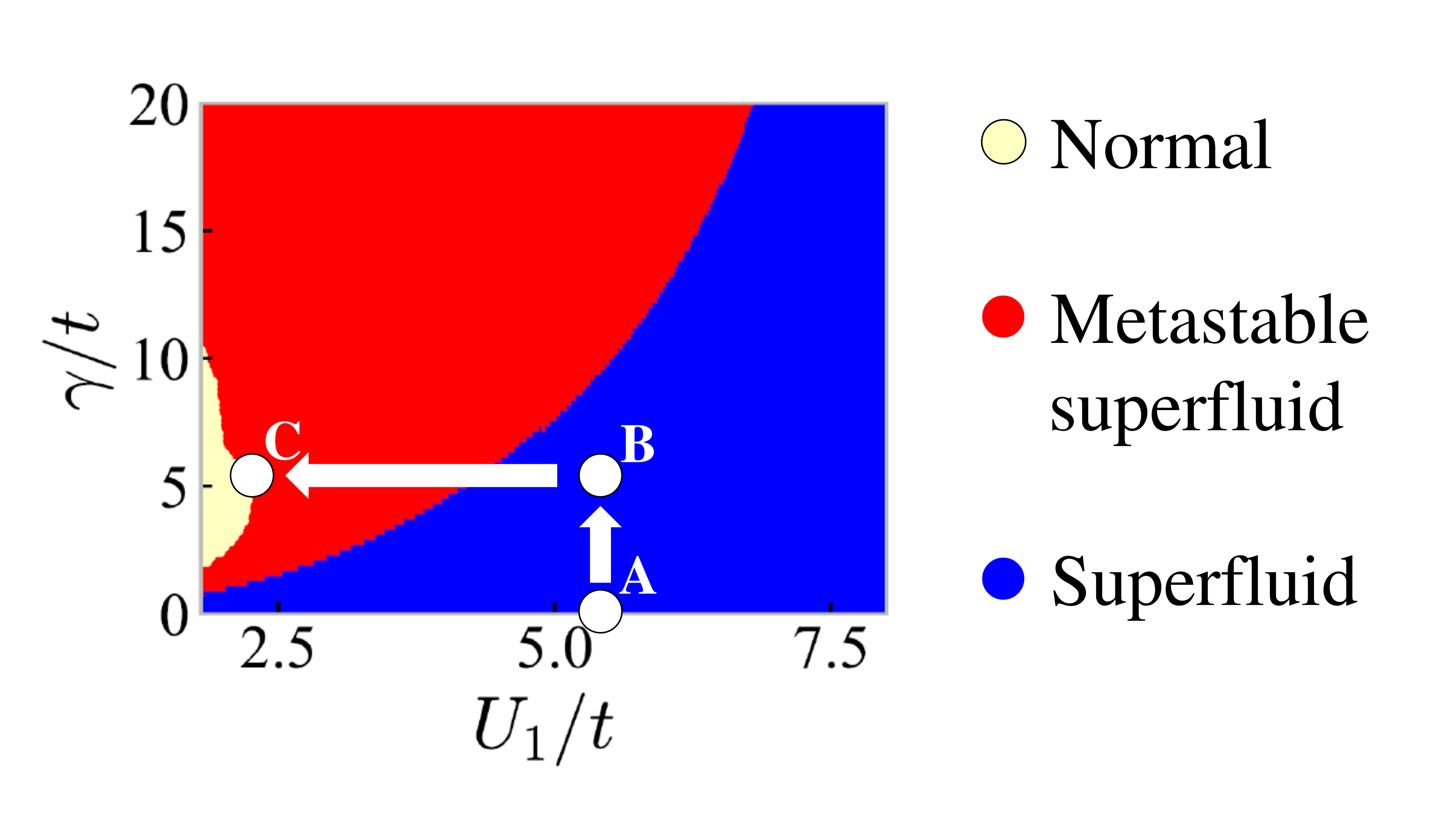}
\caption{Experimental protocol for observing the reentrant superfluidity. We first prepare on-site pairing superfluidity for large attraction $U_1$ (A) and then introduce large dissipation $\gamma$ using a sudden application of photoassocitiation (B). The system will reach the reentrant superfluid region by decreasing $U_1$ (C). }
\label{fig_experimentalrealization}
\end{figure}

The experimental protocol is illustrated in Fig.~\ref{fig_experimentalrealization}. We first prepare a superfluid state for large attraction $U_1$, where atom pairs are confined to each lattice site. Then, we introduce large dissipation $\gamma/t\sim5$ using photoassociation techniques \cite{Tomita17}. Finally, we decrease the attraction $U_1$ by tuning the magnetic field for a Feshbach resonance, which can be operated fast on the order of few $10\mu{s}$ \cite{Ketterle06}. If we do not have dissipation, atoms in the on-site molecular pairs tunnel to neighboring sites at a hopping rate $t$, which gives a timescale of the order of 100 $\mu$s \cite{Ketterle06}. However, under a large dissipation in the quantum Zeno regime, the tunneling of atoms is suppressed by the QZE and thus the dissociation of molecules after ramping down the attraction $U_1$ is delayed. Consequently, even after a timescale of $1/t$, atoms surviving in the system will still form on-site molecules. Such QZE-assisted molecules can be regarded as a signature of the reentrant superfluidity. In ultracold atoms, molecules (double occupancy) can be detected by measuring binding energies, for example, using radio-frequency spectroscopy \cite{Esslinger06}, Raman spectroscopy, or clock-laser spectroscopy \cite{Cappellini19}. Ramping down the lattice depth will give a shorter timescale of hopping, and for the region of the reentrant superfluidity, the above methods can be used to determine the phase boundaries due to exceptional points.

Finally, we note that the NH dynamics is faithfully realized when we perform a quantum-gas-microscopy measurement of an atom number and then postselect the measurement outcome which does not contain any atom loss \cite{Ashida16, Ashida17}. Since the quantum-gas microscopy for the attractive Hubbard model was already realized \cite{Bakr18}, this method can also be used for an unambiguous observation of NH superfluidity.

\end{document}